\documentclass[manuscript]{aastex}

\usepackage{natbib}
\usepackage{amsmath} 

\begin{document}

\title{Short-period pulsar oscillations following a glitch}

\author{C. A. van Eysden  }
\affil{NORDITA, Roslagstullsbacken 23, SE-10691 Stockholm, Sweden}

\begin{abstract}

Following a glitch, the crust and magnetized plasma in the outer core of a neutron star are believed to rapidly establish a state of co-rotation within a few seconds by process analogous to classical Ekman pumping.
However, in ideal magnetohydrodynamics, a final state of co-rotation is inconsistent with conservation of energy of the system.
We demonstrate that, after the Ekman-like spin up is completed, magneto-inertial waves continue to propagate throughout the star, exciting torsional oscillations in the crust and plasma.
The crust oscillation is irregular and quasi-periodic, with a dominant frequency of the order of seconds.
Crust oscillations commence after an Alfv\'en crossing time, approximately half a minute at the magnetic pole, and are subsequently damped by the electron viscosity over approximately an hour.
In rapidly rotating stars, the magneto-inertial spectrum in the core approaches a continuum, and crust oscillations are damped by resonant absorption analogous to quasi-periodic oscillations in magnetars.
The oscillations predicted are unlikely to be observed in timing data from existing radio telescopes, but may be visible to next generation telescope arrays.

\end{abstract}


\section{Introduction}

In dynamic models of pulsars, it is almost ubiquitously assumed that the strong magnetic field couples the proton-electron plasma in the outer core to the crust on short time scales, so that the two components are rigidly locked together \citep{lod84,tak89,alp84,sed95b,pas11}.
Originally proposed by \citet{bay69a}, this assumption has been validated for the slow spin-down of the pulsar frequency produced by magnetic dipole braking \citep{eas79c}, as well as for an impulsive change in the pulsar frequency induced by a glitch \citep{eas79a}.
In the latter case, co-rotation is established by an Ekman-like process, driven by the Coriolis force in the magneto-inertial boundary layers that form immediately after the glitch.  
The spin-up time is approximately two seconds for typical pulsar parameters in an ideal magnetohydrodynamic plasma, and estimated to be a factor of thirty shorter if the protons are a type II superconductor.

Questions arise, however, when one considers the differences between the classical spin up of a viscous fluid \citep{gre63} and that in an ideal magnetized plasma.  In particular, co-rotation in classical spin up is achieved via a dissipative process (viscosity), whereas ideal magnetohydrodynamics is dissipation-less. 
In both cases, a shear gradient is established immediately after the impulsive spin up of the container in the plasma abutting the boundary.  
In ideal magnetohydrodynamics, however, the corresponding shear gradient in the magnetic field stores potential energy in the form of magnetic field line tension, which provides the restoring force for Alfv\'en waves that subsequently propagate through the plasma.  
Because the plasma is dissipation-less, these waves are continually reflected internally by the container walls, generating torsional oscillations between the container and plasma. 
Oscillations of this nature have received significant attention recently in attempts to explain the quasi-periodic oscillation modes present in magnetar giant flares \citep{gla06b,lev07,sot08,col09,cer09,col11,gab11,van11c}, and earlier in the context of neutron star precession \citep{lev04}.
An argument for oscillations can be made purely on the grounds of conservation of angular momentum and energy; a final state of uniform co-rotation of the fluid and its container is not consistent with energy conservation and can {\it only} be achieved if there is dissipation.
The question arises then, why does the result of \citet{eas79a} suggest a final state of co-rotation, and what happened to the torsional oscillations of the fluid and container?

In this paper, we re-visit the problem of the spin-up of a rapidly rotating magnetized plasma after the impulsive acceleration of its container.  
We consider the same cylindrical geometry studied by \citet{eas79a}, but solve self-consistently for the motion of the plasma {\it and} its container.
We derive a general solution and compare our results with those of \citet{eas79a}, who considered only the limit in which the ratio of the rotational period to the Alfv\'en crossing time approaches zero.
We find that torsional oscillations commence after an Alfv\'en crossing time, $\sim 30\,{\rm sec}$ in a typical neutron star, and are not seen in \citet{eas79a} because the Alfv\'en crossing time is assumed to be infinitely long.
The crust oscillation amplitude also diminishes with increasing Alfv\'en crossing time because the spectrum of magneto-inertial oscillations in the core approaches a continuum, where the crust motion is damped by resonant absorption in the core analogous to magnetar oscillations \citep{lev06,lev07}.

There are two primary motivations for this study.
First, to identify and characterize the oscillatory behavior predicted above.
In the case of magnetars, extensive effort has been devoted into determining the coupled crust-core oscillation spectrum for a spherical star with realistic magnetic field configurations using numerical codes \citep{col11,gab11,van12} and analytic approaches \citep{lev07,van11c}.  The effects of a superfluid core have also been investigated \citep{and09,gab13}.
However, in the present paper we are concerned with rotation-powered pulsars, where, in contrast with magnetars, the rotational inertia dominates the magnetic forces.
Our second motivation is to determine the motion of the crust in a self-consistent manner, so that a direct comparison with radio timing data can be made, as in  \citet{van10}.
The identification and characterization of oscillations in glitch recovery could provide another dimension for using glitches to constrain properties of the pulsar interior.  

The paper is structured as follows:
The assumptions of the model and the governing equations are presented in \S\ref{sec2a}, and the solution to the system is presented in \S\ref{sec2b}.
In \S\ref{sec3a} the generic properties of the solution are explored, while in \S\ref{sec3b} the regime relevant to neutron stars is presented and the observational consequences discussed.
The conclusions are presented in \S\ref{sec4}.

\section{Coupled response of a magnetised plasma and its container} \label{sec2}

(check) Glitches are tiny, impulsive increases in the spin frequency of a pulsar.
A popular explanation is that they arise when the neutron superfluid, whose angular momentum is frozen by the pinning of superfluid vortices, is suddenly transferred to the crust as a result of avalanche unpinning of the vortex array \citep{and75b,mel09,war08,has12,war11,war12,war13}.
Following the glitch, the crust and core readjust via viscous and magnetic forces \citep{bay69a}, the latter of which is expected to be dominant in all but the oldest pulsars \citep{eas79a}.

\subsection{Governing equations} \label{sec2a}

To study the coupled response of a magnetized plasma and its container, we consider a cylindrical vessel of radius $R$ and height $2L$.  Cylindrical geometry has a long history in the study of the spin up of rapidly rotating fluids in geophysics \citep{gre63,ped67,wal69,lop71a}, condensed matter \citep{rei93} and astrophysics \citep{eas79a,abn96}.
Differences arising from spherical geometry, like that considered in \citet{van10,van13,van14} or from more realistic magnetic field geometries are discussed in \S\ref{sec3b} and are left for future studies.

The contained fluid is an incompressible magnetohydrodynamic plasma and both the container and fluid are perfect conductors.
At times $t<0$, the plasma and its container rotate rigidly and uniformly about the cylindrical axis with angular velocity $\Omega$.  The container and plasma are threaded with a uniform magnetic field of magnitude $B_0$, which is taken as  aligned with the rotation axis to keep the problem analytically tractable.  At time $t=0$ the magnitude of the angular velocity of the container is impulsively increased to $\Omega(1+\epsilon)$, where the Rossby number $\epsilon \ll 1$ in accordance with the observed sizes of pulsar glitches \citep{esp11}. For $t>0$ the container and fluid are left to evolve freely, and we solve self-consistently for the coupled motion of the container and plasma as in previous studies for viscous fluids \citep{van13,van14}.

These initial conditions model the superfluid unpinning theory of pulsar glitches described above, but can also be applied to other glitch models.  For example, if the glitch originated in the core so that the proton-electron plasma in the core super-rotated the crust, then the magnetic coupling of the crust and core considered here could model the observed spin-up of the crust during a glitch.  This scenario would correspond to $\epsilon<0$.

The magnetohydrodynamic equations for the plasma are
\begin{eqnarray}
  \partial_t  {\boldsymbol v}+{\boldsymbol v}\cdot \nabla {\boldsymbol v}&=& \frac{1}{\rho} \nabla \cdot {\bf T}\,, \label{eq1a} \\
\frac{ \partial  {\boldsymbol B}}{\partial t}&=&\nabla\times\left({\boldsymbol v}\times {\boldsymbol B}\right)\,,  \label{eq1b}\\
\nabla \cdot {\boldsymbol v} &=&0\,, \label{eq1c} \\
\nabla \cdot {\boldsymbol B} &=&0\,, \label{eq1d}
\end{eqnarray}
where ${\boldsymbol v}$ is the fluid velocity, ${\boldsymbol B}$ is the magnetic field and $\rho$ is the fluid density.
The magnetic diffusivity is omitted in (\ref{eq1b}) as it is expected to be extremely small in neutron stars \citep{eas79a}.
The stress tensor ${\bf T}$ is (in index notation)
\begin{equation}
 T_{ij}=-p \delta_{ij}+ \frac{1}{4 \pi}\left( B_i B_j -\frac{B_k B_k}{2}\delta_{ij}\right)+ \mu \left(\nabla_i v_j +\nabla_j  v_i\right) \,,\label{eq3}
\end{equation}
where $p$ is the pressure, $\mu$ is the (constant) shear viscosity, and we have employed cgs units.
The boundary conditions for the fluid are
\begin{eqnarray}
  {\boldsymbol v}={\boldsymbol \Omega}_c \times {\boldsymbol x} \,, \label{eq6}
\end{eqnarray}
at the walls of the container, where ${\boldsymbol x}$ is the radial vector and ${\boldsymbol \Omega_c}(t)$ is the angular velocity of the container, which is a function of time.  
Equation (\ref{eq6}) embodies the usual no-slip boundary conditions for viscous flows.
The boundary conditions for an inviscid magnetized plasma in a perfectly conducting container are also given by (\ref{eq6}) but only where magnetic field lines intersect the container; elsewhere only no penetration applies. 
This is discussed in more detail below.
To solve for the motion of the container, we also require the torque balance
\begin{equation}
  I_c \frac{d {\boldsymbol \Omega}_c}{ dt}=-\oint {\boldsymbol x}\times \left(\hat{\boldsymbol n} \cdot {\bf T} \right) {\rm dS}+{\boldsymbol \tau }_{ext} \,, \label{eq8}
\end{equation}
where $I_c$ is the moment of inertia of the container, $\hat{\boldsymbol n}$ is the unit vector normal to the boundary and ${\rm dS}$ is an element of area on the boundary.  The first term on the right hand side of (\ref{eq8}) is the sum of the hydrodynamic and magnetic torque exerted on the crust by the fluid and the second is a constant external torque, which may arise from e.g., the magnetic dipole torque.
In cylindrical coordinates $(r,\phi,z)$, the initial conditions are
\begin{eqnarray}
  {\boldsymbol v}(0)=r \Omega \hat{\phi}\,, \hspace{1cm}{\boldsymbol B}(0)=B_0 \hat{z}\,, \hspace{1cm} {\boldsymbol \Omega}_c(0)=\Omega(1+\epsilon)\hat{z} \,.
\end{eqnarray}

For $\epsilon \ll 1$, equations (\ref{eq1a})--(\ref{eq8}) can be linearized by perturbing around an equilibrium rotating with uniform angular velocity $\Omega$ about the $z$-axis.  
The external torque is also taken to be aligned with the rotation axis, i.e., ${\boldsymbol \tau}_{ext}=\tau_{ext} \hat{z}$.
The geometry and initial conditions are axisymmetric, and the resulting flow axisymmetric.
The following substitutions are made for the velocity, magnetic field and pressure
\begin{eqnarray}
  {\boldsymbol v} (r^*,z^*,t^*)&\rightarrow& r \Omega \hat{\phi} +  \epsilon \Omega L \left[ \frac{ \partial \psi}{\partial z^*}\hat{r}+v_\phi\hat{\phi}- \frac{1}{r^*}\frac{\partial}{\partial r^*}\left(r^*  \psi\right)\hat{z}\right]  \,, \nonumber  \\
  {\boldsymbol B} (r^*,z^*,t^*)&\rightarrow& B_0 \hat{z} +\epsilon B_0 \left[ \frac{ \partial A}{\partial z^*}\hat{r}+ B_\phi\hat{\phi}-\frac{1}{r^*}\frac{\partial}{\partial r^*}\left(r^*  A \right)\hat{z}\right] \,, \nonumber  \\
 p(r^*,z^*,t^*)&\rightarrow &  \frac{\rho \left( r^* L \Omega \right) ^2}{2}  + \frac{\epsilon \rho \left(L \Omega\right)^2}{2} P \label {eq17} \,, \label{eq9a}
\end{eqnarray}
where $\psi$, $v_\phi$, $A$, $B_\phi$ and $P$ are all dimensionless quantities and are functions of $r^*,z^*,t^*$.
The asterisked quantities are defined as $r^*=r/L$, $z^*=z/ L$, $t^*=\Omega t$. 
The form of (\ref{eq9a}) satisfies (\ref{eq1c}) and (\ref{eq1d}), and the conditions for rotational equilibrium for the background flow.
Under these assumptions, it can easily be shown that the only non-vanishing component of the external torque is in the $\hat{z}$ direction, hence
\begin{eqnarray}
 {\boldsymbol \Omega}_c(t^*)&\rightarrow & \Omega \hat{z} + \epsilon \Omega f(t^*) \hat{z} \label {eq9b} \,,
\end{eqnarray}
where $f$ is a function of $t^*$ only.

Henceforth we drop the asterisks notation, and all coordinates refer to the dimensionless variables.
Substituting (\ref{eq9a}) into the induction equation (\ref{eq1b}), we obtain
\begin{eqnarray}
  \frac{\partial B_\phi }{\partial t} &=&  \frac{ \partial v_{\phi}}{\partial z} \,, \label{eq10} \\
 \frac{\partial A }{\partial t} &=&  \frac{ \partial \psi}{\partial z} \,.  \label{eq11}
\end{eqnarray}
Substituting (\ref{eq9a}) into the momentum equation (\ref{eq1a}) and eliminating $P$, we obtain
\begin{eqnarray}
\left(\frac{\partial }{\partial t} -E \mathcal{L} \right) \mathcal{L}\psi &=&2 \frac{\partial v_{\phi}}{\partial z}  +  \xi^2 \frac{\partial}{\partial z} \mathcal{L} A \,,  \label{eq12} \\
\left(\frac{\partial }{\partial t} -E \mathcal{L} \right) v_\phi &=&-2  \frac{\partial \psi}{\partial z}+\xi^2\frac{ \partial }{\partial z} B_\phi  \,,  \label{eq13a}
\end{eqnarray}
where we define
\begin{eqnarray}
\mathcal{L}= \frac{\partial}{\partial r}\frac{1}{r}\frac{\partial}{\partial r} r+\frac{\partial^2}{ \partial z^2} \,,
\end{eqnarray}
and the dimensionless parameters
\begin{eqnarray}
E=\frac{\mu}{\rho L^2 \Omega} \,,\hspace{1cm} \xi=\frac{1}{L\Omega}\sqrt{\frac{B_0^2}{4 \pi \rho}}\,. \label{eq16}
\end{eqnarray}
The first parameter in (\ref{eq16}) is commonly referred to as the Ekman number and is a measure of the ratio of the viscous forces to the inertial forces.
The second is the ratio of magnetic forces to inertial forces, and is proportional to the ratio of the rotational period and the Alfv\'en crossing time.
Substituting (\ref{eq9a}) and (\ref{eq9b}) into (\ref{eq8}), we obtain 
\begin{eqnarray}
 \frac{d f}{dt}&=&-4 K \int_0^1 \left(\xi^2 B_\phi+E\frac{\partial v_\phi}{\partial z} \right)_{z=1} r^2 dr \nonumber \\
 &&-4 K E \int_{0}^1 \left[r^3  \frac{\partial }{\partial r} \left( \frac{ v_\phi}{r} \right)\right]_{r=R/L}dz+\alpha \left(1+K\right)\,. \label{eq17}
\end{eqnarray}
where we define
\begin{equation}
  K=\frac{ \pi \rho R^4 L}{I_c}\,, \hspace{1cm}  \alpha=\frac{\tau_{ext}}{\epsilon \Omega^2 I_{tot}}\,, \label{eq18}
\end{equation}
and we have utilized the symmetry of the container about the $z=0$ plane to obtain (\ref{eq17}).
In (\ref{eq18}), the dimensionless parameter $K$ denotes the ratio of the moments of inertia of the fluid and container, and $\alpha$ is the dimensionless external torque where $I_{tot}=I_c (1+K)$ is the total moment of inertia of the fluid and container.

The first term and second terms on the right-hand side of (\ref{eq17}) are the contributions to the torque from the horizontal and vertical boundaries respectively.
Note that the magnetic field does not contribute to the torque on the vertical boundaries of the container. 
The reason for this is connected with the boundary conditions of the container, which we examine now.
Combining (\ref{eq10})--(\ref{eq13a}) we obtain the governing equation
\begin{equation}
  \left[ \left(\frac{\partial^2 }{\partial t^2} -E \frac{\partial }{\partial t} \mathcal{L} -\xi^2 \frac{\partial^2}{\partial z^2}\right)^2\mathcal{L}+4 \frac{\partial^2}{\partial z^2} \frac{\partial^2 }{\partial t^2} \right]v_\phi=0\,. \label{eq15}
\end{equation}
When the viscosity is non-zero, (\ref{eq15}) is a sixth order partial differential equation in $r$ and $z$.  
The boundary conditions required for a unique solution are obtained from (\ref{eq6}) and (\ref{eq9a}), namely
\begin{eqnarray}
 &&v_\phi=r f\,,  \frac{ \partial \psi}{\partial z}=0\,, \frac{\partial}{\partial r}\left(r  \psi\right)=0\,, {\rm at}\; z=\pm 1\,, \nonumber \\
 &&v_\phi=R f/L \,,  \frac{ \partial \psi}{\partial z}=0\,, \frac{\partial}{\partial r}\left(r  \psi\right)=0\,, {\rm at}\; r=1\,, \nonumber \\
 &&v_\phi=0\,,  \frac{ \partial \psi}{\partial z}=0\,, \frac{\partial}{\partial r}\left(r  \psi\right)\,{\rm finite\,at}\; r=0\,.
\end{eqnarray}
which are the usual no-slip boundary conditions for viscous flow.
However, in the absence of viscosity ($E=0$), (\ref{eq15}) is sixth order in $z$, but only second order in $r$.  
Clearly, in this case there is not enough freedom to apply all the boundary conditions above.
This is a consequence of the geometry of the magnetic field.
At the horizontal boundaries, the magnetic field lines intersect the container.
Because the plasma and container are perfect conductors, flux freezing requires that the velocity of the fluid must always match the velocity of the container, i.e., no-slip.
Thus, a change in velocity of the container is communicated to the fluid via the magnetic field.
However, at the vertical boundaries, the magnetic field lines do not intersect the container.
Therefore, there is no requirement that the velocities of the container and fluid match, and no torque is exerted on the vertical boundaries.
The required boundary conditions in this case are
\begin{eqnarray}
 &&v_\phi=r f\,,  \frac{ \partial \psi}{\partial z}=0\,, \frac{\partial}{\partial r}\left(r  \psi\right)=0\,, {\rm at}\; z=\pm 1\,, \nonumber \\
 && \frac{ \partial \psi}{\partial z}=0\,, {\rm at}\; r=1\,, \nonumber \\
 && \psi=0\,,\, {\rm \,at}\; r=0\,. \label{eq23}
\end{eqnarray}
Hence, {\it only} no-penetration is applied at the vertical boundaries when there is no viscosity.
Note that the most general magneto-hydrodynamic equations describing neutron star cores are more complicated that those considered here \citep{gla11}, and this set of equations requires a more thorough consideration of the boundary conditions.
Finally, in terms of the perturbed quantities, the initial conditions are
\begin{equation}
 v_\phi=\psi =B_\phi=A=0\,, f=1\,,\; {\rm at} \, t=0\,.
\end{equation}

\subsection{Solution} \label{sec2b}

To solve (\ref{eq10})--(\ref{eq18}), we follow previous authors \citep{gre63,eas79a} and consider solutions of the form 
\begin{equation}
 v_\phi=r V (z)\,,\;\; \psi=r \chi (z)\,. \label {eq27}
\end{equation}
The boundary conditions become
\begin{eqnarray}
 &&V=f\,,  \frac{ \partial \chi}{\partial z}=0\,, \chi=0\,, {\rm at}\; z=\pm 1\,, \label{eq28}
\end{eqnarray}
for all values of $\xi$ and $E$.
Note that adopting the form (\ref{eq27}), we no longer impose the boundary conditions at $r=1$.
A solution of this form is extensively used for spin-up problems in infinite parallel plate geometry, i.e., containers of infinite aspect ratio $(R/L)$.
However, in addition to containers of infinite aspect ratio, this solution also applies to containers of arbitrary aspect ratio when rapidly rotating with $\xi \ll 1$ and $E \ll 1$, where the Stewartson layers form to satisfy the boundary conditions at the vertical boundaries \citep{gre68}.
This is discussed in more detail in  \S\ref{sec3}.

To solve the system, we employ the Laplace transform 
\begin{equation}
  \tilde{X}(z,s)=\int_0^\infty X(z,t) e^{-s t} {\rm d} t\,.
\end{equation}
Using (\ref{eq10}) and (\ref{eq11}) to eliminate $B_\phi$ and $A$, (\ref{eq12}), (\ref{eq13a}) and (\ref{eq17}) become
\begin{eqnarray}
\left[s^2 -\left( E s + \xi^2\right) \frac{\partial^2}{\partial z^2}  \right] \frac{\partial^2  \tilde{\chi} }{\partial z^2 } &=&2 s \frac{\partial \tilde{V}}{\partial z}    \,,  \label{eq12a} \\
\left[s^2- \left(E s+ \xi^2\right)\frac{ \partial^2 }{\partial z^2} \right] \tilde{V} &=&-2 s \frac{\partial \tilde{\chi}}{\partial z} \,,  \label{eq13b} \\
s\left(s \tilde{f}-1\right)&=&- K \left(\xi^2 +E s  \right) \left( \frac{\partial \tilde{V}}{\partial z} \right)_{z=1}+\alpha (1+K)\,,
\end{eqnarray}
which must be solved subject to the boundary conditions (\ref{eq28}).
The solution is 
\begin{eqnarray}
  \tilde{V}&=& \tilde{f} \left[i C_+ \left( \cosh k_+ z-\cosh k_+ \right)- i C_- \left( \cosh k_- z-\cosh k_- \right) +1 \right]\,,  \label{eq31} \\
 \tilde{\chi}&=& - \tilde{f} \left[\frac{C_+}{k_+ } \left(  \sinh k_+ z-z \sinh k_+ \right)+\frac{C_-}{k_- } \left(\sinh k_- z-z \sinh k_- \right)\right]\,. \\
 \tilde{f} &=& \left[\frac{1}{s}+\frac{\alpha\left(1+K\right)}{s^2}\right]  \frac{ \Delta }{ \bar{\Delta} }  \label{eq33}
\end{eqnarray}
where
\begin{eqnarray}
k_\pm&=&\sqrt{\frac{s^2 \pm i 2  s}{E s+\xi^2}}\,, \\
 C_\pm&=&\frac{D_\mp k_\pm  s \left(k^2_\mp -k_\pm^2\right)}{4  \Delta }\,, \\
  D_\pm&=& k_\pm  \cosh k_\pm -  \sinh k_\pm \,, \\
\Delta&=&k_-^{3} D_+\cosh k_-+k_+^3 D_- \cosh k_+\,, \\
  \bar{\Delta}&=&k^{2}_- D_+ \left( k_-  \cosh k_- +K \sinh k_- \right)  + k^2_+ D_- \left( k_+ \cosh k_++K \sinh k_+ \right) \,.
\end{eqnarray}
The above results reduce to the expression obtained by \citet{gre63} in the limit $\xi=K=\alpha=0$.  The results of the elegant and compact notation of \citet{eas79a} are also recovered in the limit $K=\alpha=0$; by expanding the real and complex components of his complex function $F$ one obtains $F=\partial \chi/\partial z+i V$.

The motion of the crust is the observable quantity of interest, therefore require the inverse Laplace transform of (\ref{eq33}).
A cursory examination reveals that when $\alpha=0$, all the poles are simple poles, located at $s=0$ and at the zeros of $\bar{\Delta}$.
By observing that the terms multiplied by $\alpha$ are equal to the expression for $\alpha=0$ divided by $s$, we deduce that the inverse Laplace transform of the terms multiplied by $\alpha$ are simply the integral of the inverse Laplace transform for $\alpha=0$.
The complete solution is
\begin{equation}
  f(t)=\frac{1}{1+K}+\alpha t+\sum_n R(s_n) \left[e^{s_n t}+\frac{\alpha(1+K)}{s_n}\left(e^{s_n t}-1\right)\right] \,,\label{eq39}
\end{equation}
where the first term arises from the poles at $s=0$ and the sum arises from the poles corresponding to the zeroes of $\bar{\Delta}$, denoted $s_n$.  
The coefficients $R(s_n)$ are obtained from the residue calculation and are given by
\begin{equation}
  R(s) = \frac{\Delta (s)}{s \frac{d \bar{\Delta}}{d s}}\,.
\end{equation}
Equation (\ref{eq39}) is a general solution for the coupled evolution of a container of infinite aspect ratio and the contained plasma, valid for any $K$, $E$, and $\xi$.  
Previous studies consider the limit $E \ll 1$, $\xi \ll 1$ and $s\ll 1$ to invert the Laplace transform \citep{gre63,eas79a}, for which an elegant analytical solution is obtained that elucidates the physics of Ekman pumping.
However, this solution {\it only} resolves the physics of Ekman pumping and filters other aspects of the problem such as short time-scale oscillations.
The solution (\ref{eq39}) facilitates a {\it complete} investigation of the system, and we demonstrate that some of the physics that is neglected in \citet{eas79a} is relevant to glitch recovery.

Useful analytic expressions for the inverse Laplace transform of (\ref{eq33}) can be obtained in two limits: fast rotation and slow rotation.  In the slow rotation limit $\Omega\approx 0$ and disappears from the governing equations. An exact solution can be obtained for the inverse Laplace transform and is given in \S\ref{secA1}.  In the fast rotation limit, presented in \S\ref{secA2}, we have either $\xi \ll 1$ or $E \ll 1$, where generalizations of the solutions obtained by \citet{eas79a} and \citet{gre63} respectively are obtained for arbitrary $K$.

\section{Results and discussion} \label{sec3}

\subsection{Features of the solution} \label{sec3a}

To evaluate (\ref{eq39}), the major task is obtaining the required zeros of $\bar{\Delta}$ and evaluating the $R(s_n)$.  This can be readily done using, e.g., Mathematica.  
In general, the zeros and their corresponding residues are either real or occur in complex conjugates and their values depend on $K$, $E$ and $\xi$.  
For $\alpha=0$, (\ref{eq39}) can be written
\begin{equation}
  f(t)=\frac{1}{1+K}+2 \sum_n e^{{\rm Re} (s_n) t} \left\{ {\rm Re} [R(s_n)]  \cos[{\rm Im}(s_n) t] - {\rm Im} [R(s_n)]  \sin[{\rm Im}(s_n) t]\right\} \,, \label{eq41a}
\end{equation}
where the sum now refers to zeros of $\bar{\Delta}$ with a positive imaginary component.
Note that the zeros of $\bar{\Delta} $ at $s=\pm 2 i$ are also a zero in the numerator, $\Delta$,  and hence do not correspond to poles of $\tilde{f}$.
One finds that always ${\rm Re} (s_n)\leq 0$, and vanishes when $E=0$ (along with Im$[R(s_n)]$).
Significant care must be taken to ensure that one has obtained enough terms for convergence of the solution, and that one has thoroughly searched the complex plane for all significant contributions.  The number of terms required strongly depends on $E$ and $\xi$, and the convergence of the solution is discussed for particular examples below.

\begin{figure}
\epsscale{0.75}
\plotone{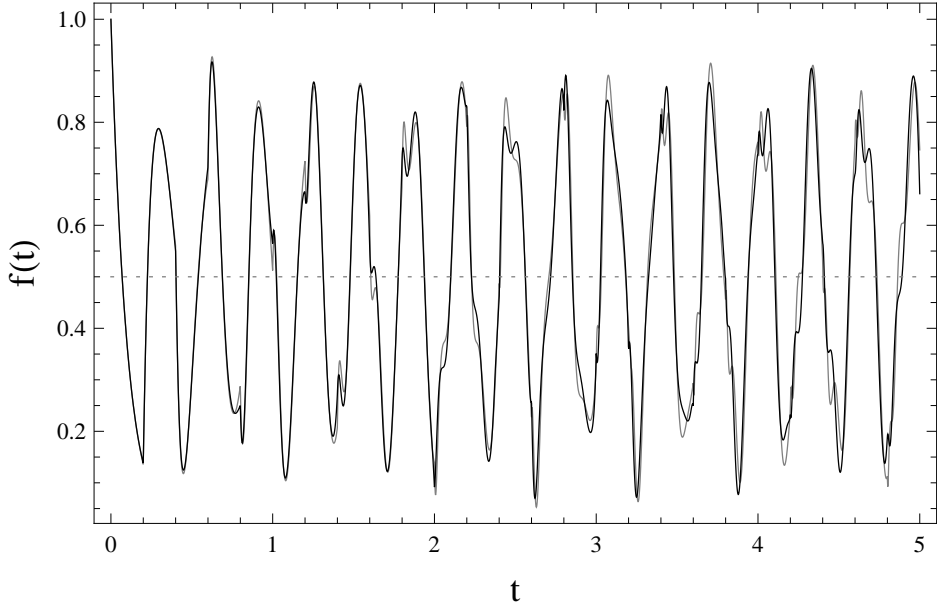}
\caption{Motion of the container, $f(t)$, for $\xi=10$, $K=1$, $E=0$ and $\alpha=0$. The black curve corresponds to the exact solution (\ref{eq41a}), while the gray curve corresponds to the asymptotic solution in the slow rotation limit (\ref{eqA3}).  The dotted line corresponds to the centre-of-mass of the system $1/(1+K)$. \label{fig1}}
\end{figure}

\begin{figure}
\epsscale{0.40}
\plotone{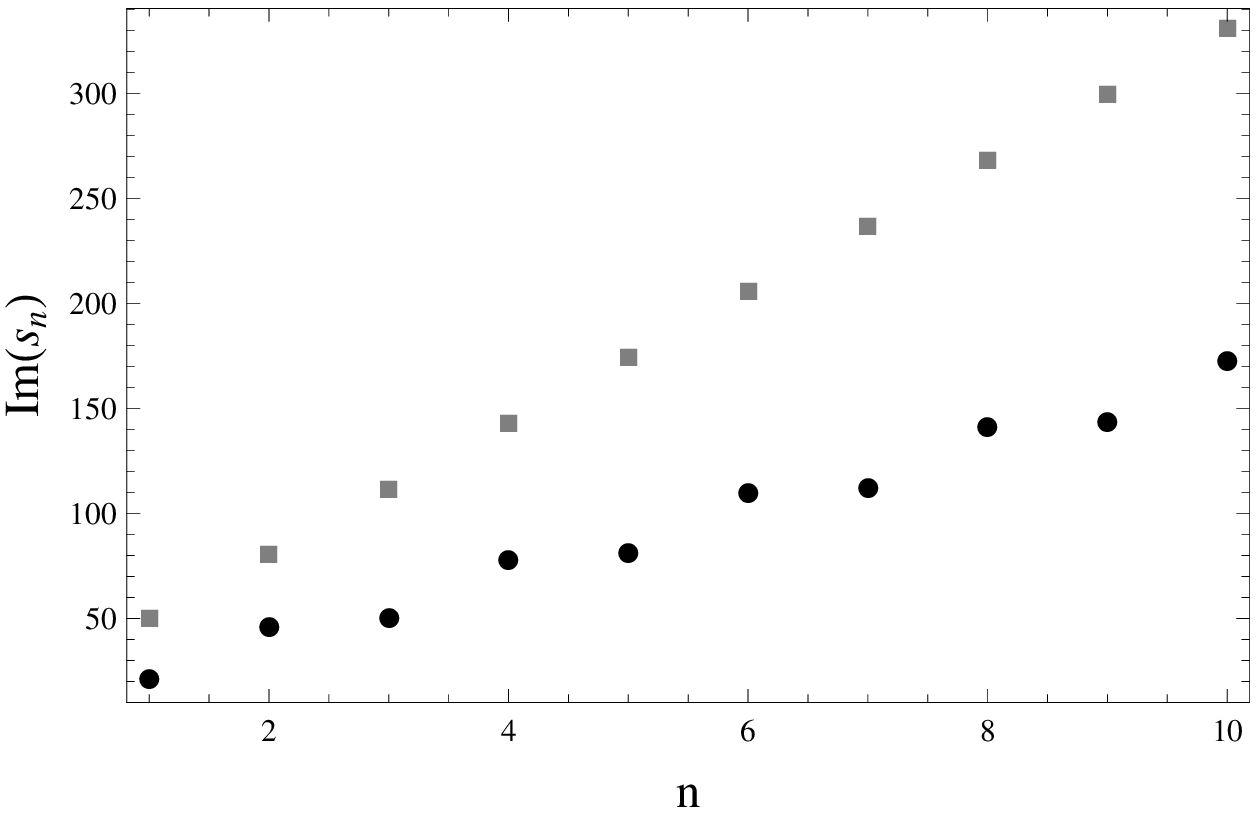} 
\plotone{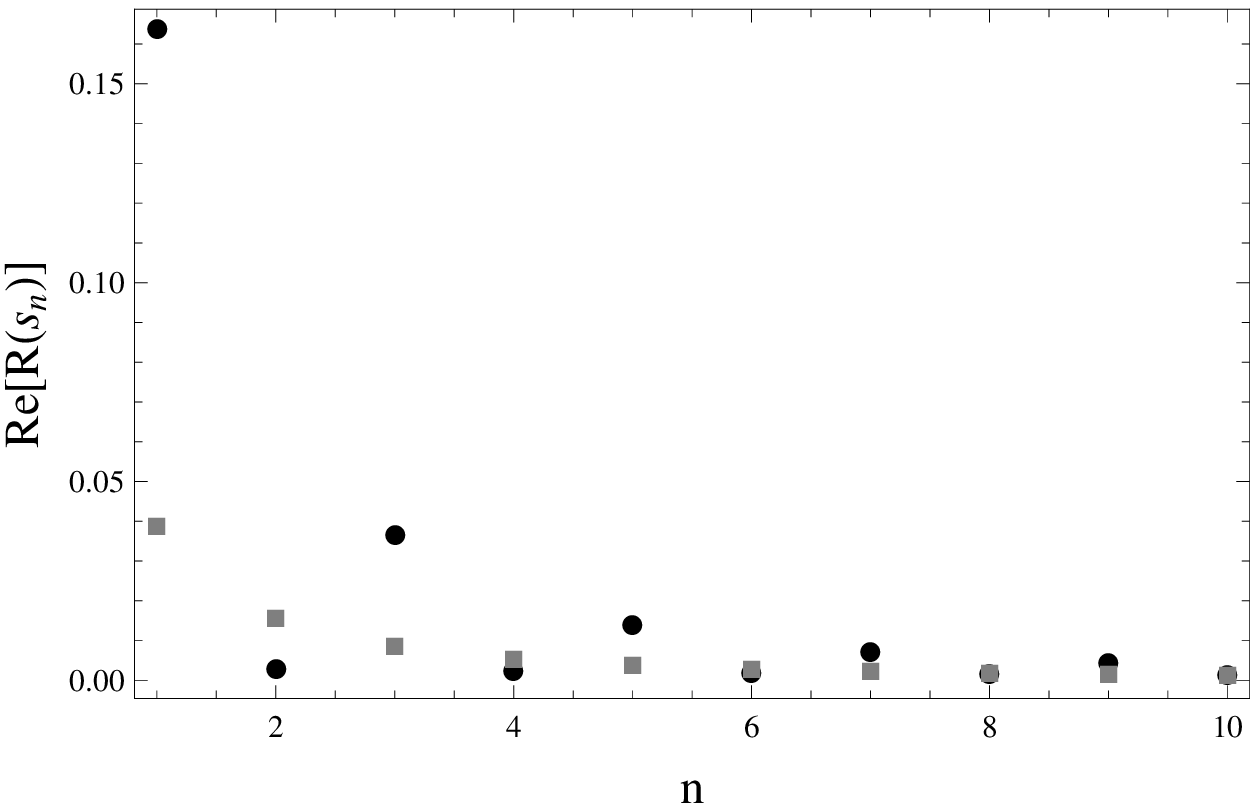} 
\caption{The imaginary component of the eigenvalues $s_n$ (left-hand panel) and the real component of $R(s)$ (right-hand panel) for the motion of the container in figure \ref{fig1}.  The real component of $s_n$ and imaginary component of $R(s_n)$ vanish for $E=0$.  The black circles correspond to the exact solution, while the gray squares corresponds to the asymptotic solution.\label{fig2}  }
\end{figure}

To begin our study, let us consider the simplest case of the interaction between the magnetized plasma and its container.  We consider the limit $\xi \gg 1$, $E=0$, $\alpha=0$, i.e., the rotational period is much longer than the Alfv\'en crossing time, the fluid is inviscid and there is no external torque.  In this limit, the poles of $\bar{\Delta}$ are readily found, and an elegant analytic solution is obtained, given by (\ref{eqA3}) in \S\ref{secA1}.  Because rotational effects are negligible, the radial velocity is no longer coupled via the Coriolis force and the flow is purely azimuthal.  
Therefore, the boundary conditions at $r=R/L$ given in (\ref{eq23}) are automatically satisfied, and the solution applies to containers of finite aspect ratio.

In figure \ref{fig1}, the exact solution (\ref{eq41a}) is plotted as a black curve for $\xi=10$, $K=1$, $E=0$ and $\alpha=0$.
For comparison, the asymptotic solution for $\xi \gg 1$ given by (\ref{eqA3}) is plotted as a gray curve in the same figure, using the same parameters as the exact result.
In both cases, the first three thousand eigenmodes were included, ordered by increasing Im($s_n$).
The angular velocity of the centre of angular momentum frame, $1/(1+K)$, is plotted as a gray dotted line.
Comparing the black and gray curves, it is immediately apparent that for $\xi=10$ the exact solution closely matches the asymptotic result.  
This establishes confidence in the correctness of both results, which are obtained finding the inverse Laplace transform of (\ref{eq33}) using two independent methods.
Examining figure \ref{fig1}, we find that the solution is oscillatory, as predicted.
The impulsive increase in angular momentum of the container excites magneto-inertial waves that propagate through the plasma, exchanging angular momentum with the container as they are reflected internally.
This is fundamentally different from the analogous problem in a viscous fluid, where the system relaxes to a state of co-rotation due to dissipation [see (\ref{eqA3}) in the limit $\xi \ll E$].


The second feature present in figure \ref{fig1} is that the oscillations are irregular.  
This arises from the many degrees of freedom in the system, and should not be confused with chaos or randomness.
The motion of the container is analogous to that of a mass that is coupled to many smaller masses via springs, where the smaller masses and springs are analogous to the fluid and magnetic field line tension respectively.  
This produces a very complicated, but completely deterministic motion. 
This irregularity can also be understood by examining the eigenvalues of the system, which are unevenly spaced.
In the left-hand panel of figure \ref{fig2}, the imaginary component of the eigenvalues $s_n$ are presented; the real component vanishes for $E=0$.  The eigenvalues for the exact solution, located at the zeroes of $\bar{\Delta}$, are plotted as black circles, whereas the eigenvalues for asymptotic solution, obtained from (\ref{eqA4}) and (\ref{eqA5}), are plotted as gray squares.
Clearly, for the exact solution, the eigenvalues are unevenly spaced, where the spacing between adjacent eigenvalues alternates between small and large values, creating a staircase pattern.
Although not obvious from figure \ref{fig2}, the eigenvalues for the asymptotic solution are also unevenly spaced; the difference between adjacent eigenvalues approaches $\xi \pi$ as $n\rightarrow \infty$.

In the right-hand panel of figure \ref{fig2}, the real component of $R(s)$ is presented; the imaginary component vanishes for $E=0$.  This illustrates the relative weighting of each mode, i.e., the Fourier transform of (\ref{eq41a}).  Both the asymptotic and exact solutions are presented, as in the left-hand panel.
For the asymptotic solution, the Re[$R(s_n)$] decrease smoothly as $n$ is increased. 
For the exact solution, the first mode is dominant, and subsequent Re[$R(s_n)$] lie alternately above and below the corresponding ones for the asymptotic solution.
It can be shown that the exact solution converges to the asymptotic one as $\xi \rightarrow \infty$ in both panels of figure \ref{fig2}.  
Both the asymptotic and exact solutions for Re[$R(s_n)$] are small by $n=10$, suggesting that this is sufficient for convergence (recall that three thousand eigenmodes have been plotted).
Another useful convergence check is the initial condition $f(0)=1$.  Using 10 terms, $f(0)=0.9615$, whereas with 3000 terms, $f(0)=0.9999$.

The slow rotation limit considered in figures \ref{fig1} and \ref{fig2} ($\xi\ll1$) lies in the regime relevant for magnetars, for which the oscillation modes have been studied extensively \citep{gla06b,lev07,sot08,col09,cer09,col11,gab11,van11c}.
\citet{lev06,lev07} argued that in spherical geometry, the spectrum of Alfv\'en oscillations in the core is a continuum, and the resonant absorption of global modes by the continuum results in the damping of crust oscillations (referred to as Landau damping) in a fraction of an Alfv\'en crossing time.  This contrasts with the present result in cylindrical geometry, in which the Alfv\'en spectrum is discrete and no damping is observed.  Hence, for $\xi\ll1$, the difference between cylindrical and spherical geometry produces important, qualitatively different results.
However, persistent quasi-periodic oscillations may still exist and correspond to the turning points and edges of the continuum \citet{lev07,cer09,col09}.
The situation becomes more complicated when considering realistic magnetic field configurations \citep{col09,van12}, superfluidity in the core \citep{gab13,pas14}, and non-axisymmetric oscillation modes \citep{lan10,lan11}, which can break up the continuum.
Modes lying in the gaps of the continuum are not subject to Landau damping and correspond to persistent quasi-periodic oscillations.
Therefore, a realistic prediction for the quasi-periodic oscillations requires detailed modeling.
We do not intend to explore these features here, only establish the behavior of the system before considering the regime of interest to pulsars.

\begin{figure}
\epsscale{0.75}
\plotone{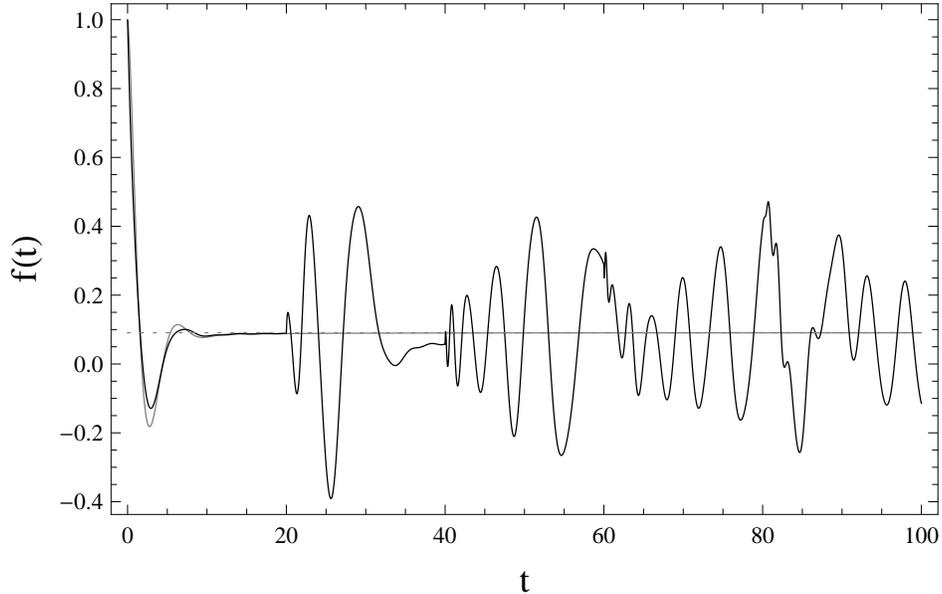}
\caption{Motion of the container, $f(t)$, for $\xi=0.1$, $K=10$, $E=0$ and $\alpha=0$. The black curve is the exact solution (\ref{eq41a}), while the gray curve corresponds to the approximate solution in the fast rotation limit (\ref{eqA12}).  The dashed line corresponds to the centre-of-mass of the system, $1/(1+K)$. \label{fig3}}
\end{figure}

\begin{figure}
\epsscale{0.40}
\plotone{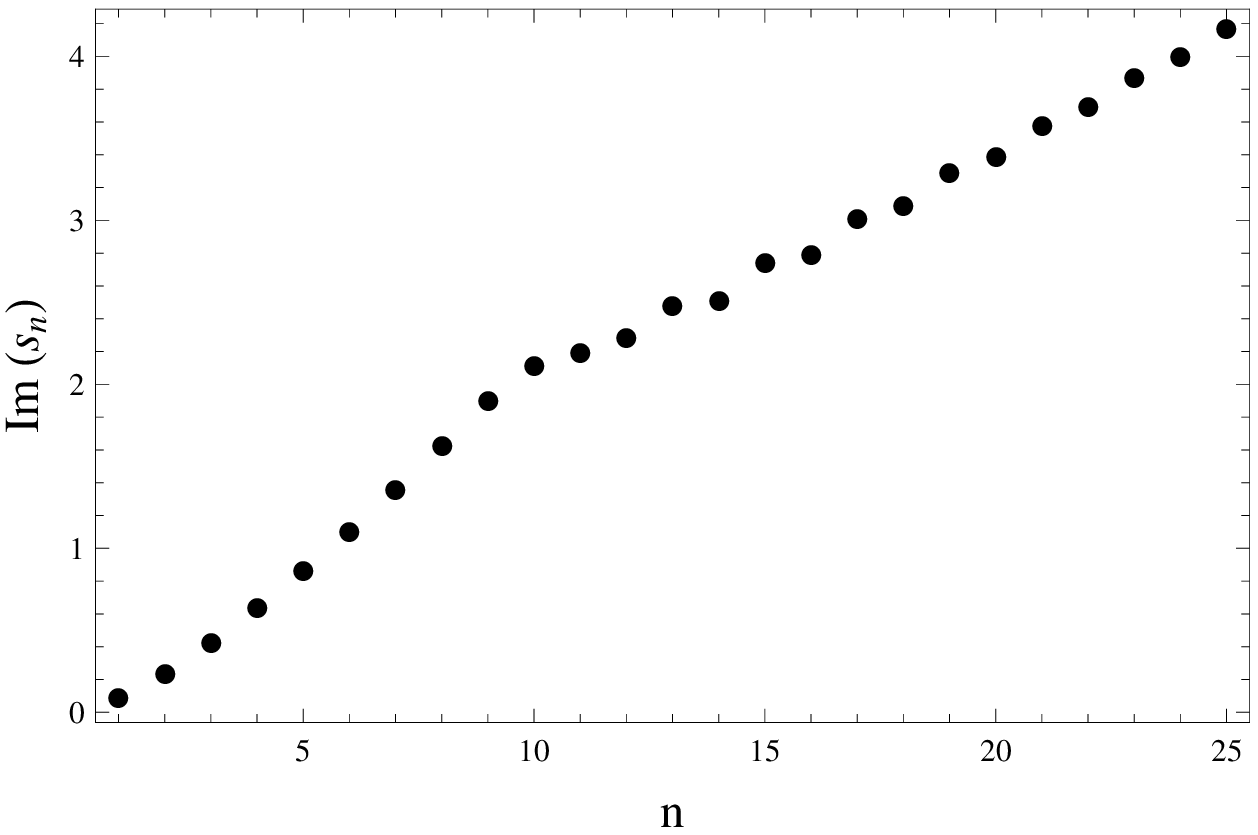} 
\plotone{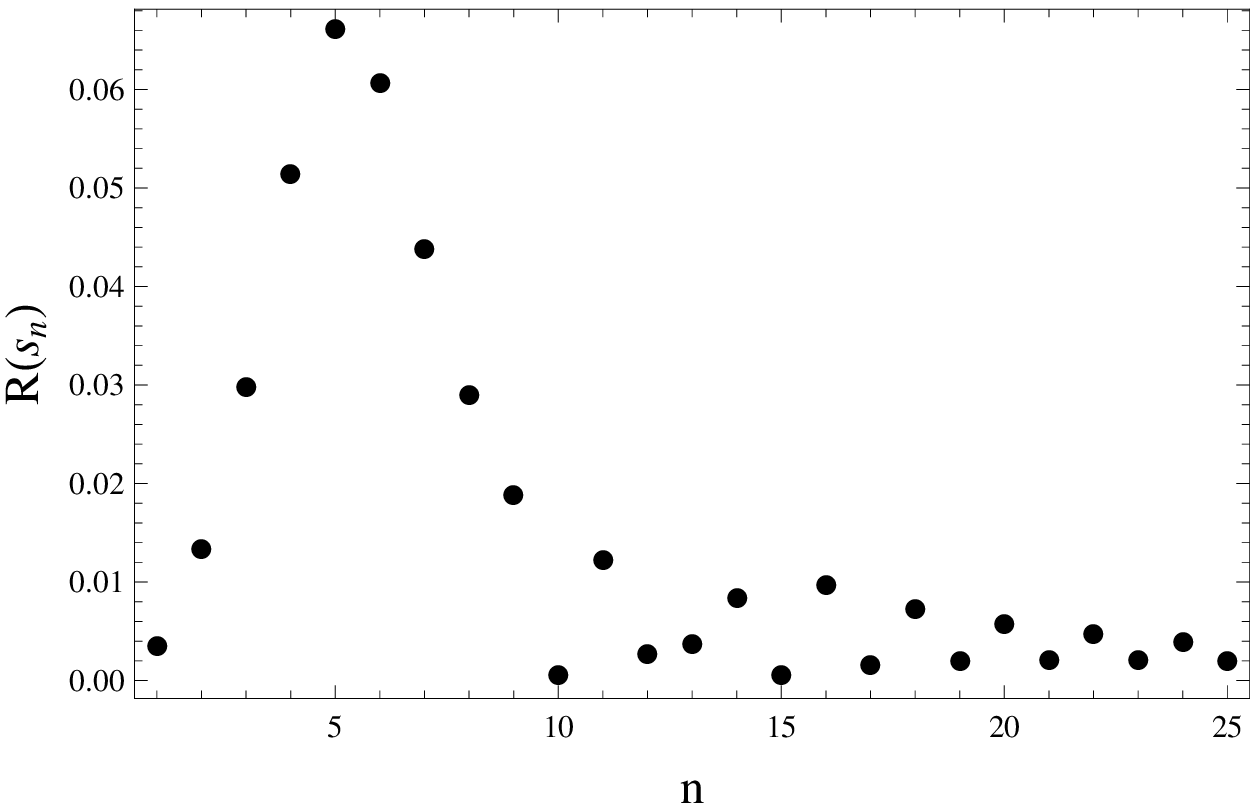} 
\caption{The imaginary component of the eigenvalues $s_n$ (left-hand panel) and the real component of $R(s_n)$ (right-hand panel) for the motion of the container in figure \ref{fig3}. \label{fig4} }
\end{figure}

Having examined the features of the solution in the limit of slow rotation ($\xi\gg 1$), we now consider rapid rotation ($\xi\ll 1$) corresponding to isolated radio pulsars.  
In figure \ref{fig3}, (\ref{eq41a}) is plotted as a black curve for $\xi=0.1$, $K=10$, $E=0$ and $\alpha=0$.
Compared with figure \ref{fig1}, the container is now much lighter and the rotational period much shorter than the Alfv\'en crossing time.  The viscosity and external torque are neglected and 3000 eigenmodes have been plotted as before. 
An asymptotic solution in the limit $\xi \rightarrow 0$ is derived in (\ref{eqA12}), and corresponds to the generalization of  \citet{eas79a} to include the self-consistent motion of the container. 
This solution is plotted as a gray line in figure \ref{fig3}.   
The gray-dotted curve represents the centre of angular momentum of the system as before.

It is instructive to study the asymptotic solution first.  
For rapidly rotating containers ($\xi\ll 1$), the magnetized plasma in the interior spins up via an Ekman-like process.  
Classical Ekman pumping occurs in three phases \citep{gre63}: (i) formation of a viscous boundary layer on the horizontal boundaries, (ii) the Coriolis force in the boundary layers drives a radial outflow, producing a secondary flow that cycles fluid through the boundary layer and back into the interior, and (iii) the damping out of residual oscillations.
The process (ii) spins up the interior fluid on a timescale $E^{-1/2} \Omega^{-1}$ (in dimensional units), known as the Ekman time.
The spin up of a magnetized plasma proceeds in an analogous manner, with a spin-up time $\xi^{-2/3} \Omega^{-1}$ (in dimensional units)\citep{eas79a}.  
The qualitative features identified by \citet{eas79a} are present; the system overshoots at $t\sim 2$, oscillates once, and then relaxes to a state of co-rotation.

Comparing the asymptotic solution with the exact solution, we see that the solutions match closely until 
$t\approx 20$, at which point the exact solution begins to oscillate.
Oscillations are expected, as demonstrated in figure \ref{fig3}, but why do they begin at $t\approx 20$ in the exact solution and not at all in the asymptotic solution?
The reason is as follows:
At $t=0$, the impulsive acceleration of the boundaries excites magneto-inertial waves that propagate into the plasma.  
For $t < \xi^{-2/3} $, these waves have only travelled a distance $ L \xi^{2/3} $ (in dimensional units) and comprise the boundary layers that facilitate Ekman pumping, which spins up the plasma into co-rotation with the container as predicted by \citet{eas79a}.
However, this is not strictly a state of co-rotation, as the magneto-inertial waves continue to propagate through the plasma after the spin-up has completed.
After a time $t=2 \xi^{-1}$, the magneto-inertial wave crosses the container (of dimensional height $2 L$), exerting a torque on the boundary as it is reflected, generating the observed oscillations.  
The magneto-inertial waves continue to be reflected internally at later times, producing oscillations like those in figure \ref{fig1}.  
Therefore the oscillations commence after an Alfv\'en crossing time, and are not seen in the analysis of \citet{eas79a} because the limit $\xi \ll1 $ is assumed where the Alfv\'en crossing time is infinitely long.

In figure \ref{fig4}, we present the eigenvalues Im($s_n$) and mode weightings Re[$R(s_n)$] in the left- and right-hand panels respectively; the remaining components vanish as in figure \ref{fig2}.
Because of the complicated dependence of the $s_n$ on $K$ and $\xi$, the eigenmodes do not follow any distinct pattern.  
The dominant mode is $s_5=0.85$, and has a corresponding period of $7.4$ in dimensionless units.
Using 10 terms, $f(0)=0.7213$, whereas after 3000 terms, $f(0)=0.9987$.
More terms are required for convergence than for $\xi=10$, as the boundary layer structure in the plasma must be resolved.

The solution presented in figure \ref{fig3} is applicable to containers of arbitrary aspect ratio.
In classic Ekman pumping, the boundary conditions at $r=R/L$ are satisfied by the formation the double-deckered  boundary layers known as the Stewartson layers, which perform the following functions:
A layer of thickness $E^{1/4}$ forms in the azimuthal velocity, which equilibrates the azimuthal velocity of the interior flow with that of the container.
A second layer of thickness $E^{1/3}$ forms that satisfies no-slip for the vertical velocity component, but also satisfies no penetration for the radial flow by recycling the $O(E^{1/2})$ flux in the Ekman layers into the interior.
These layers have no dynamic effect on Ekman pumping, but form to satisfy the required boundary conditions at the vertical boundaries.
Therefore, for $E \ll 1$, the solution for infinite parallel plates applies to containers of finite aspect ratio.
An analogous situation occurs for the spin up of a magnetized plasma.
The formation of Stewartson-like layers at the vertical boundaries in this case is less well studied, however, from (\ref{eq23}) it is obvious that the boundary layer structure will be less complicated than for the viscous case as it must only satisfy the no-penetration requirement for the radial flow.  
A detailed study of this layer is not presented here, but it suffices to reason that it recycles the $O(\xi^{2/3})$ flux from the horizontal boundary layers back into the interior, analogous to the viscous case.
Therefore the solution (\ref{eq41a}) is applicable to containers of arbitrary aspect ratio in both the limits $\xi \gg 1$ and $\xi \ll 1$.

\begin{figure}
\epsscale{0.40}
\plotone{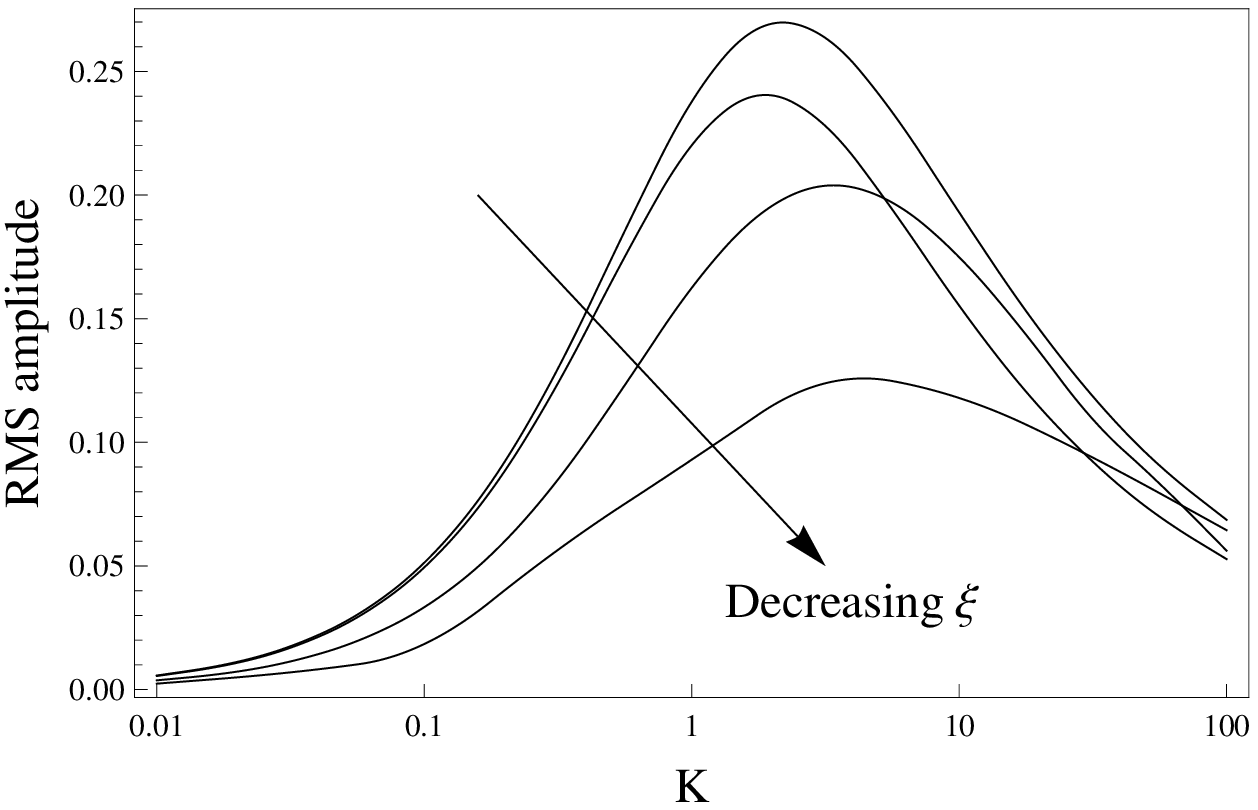} 
\plotone{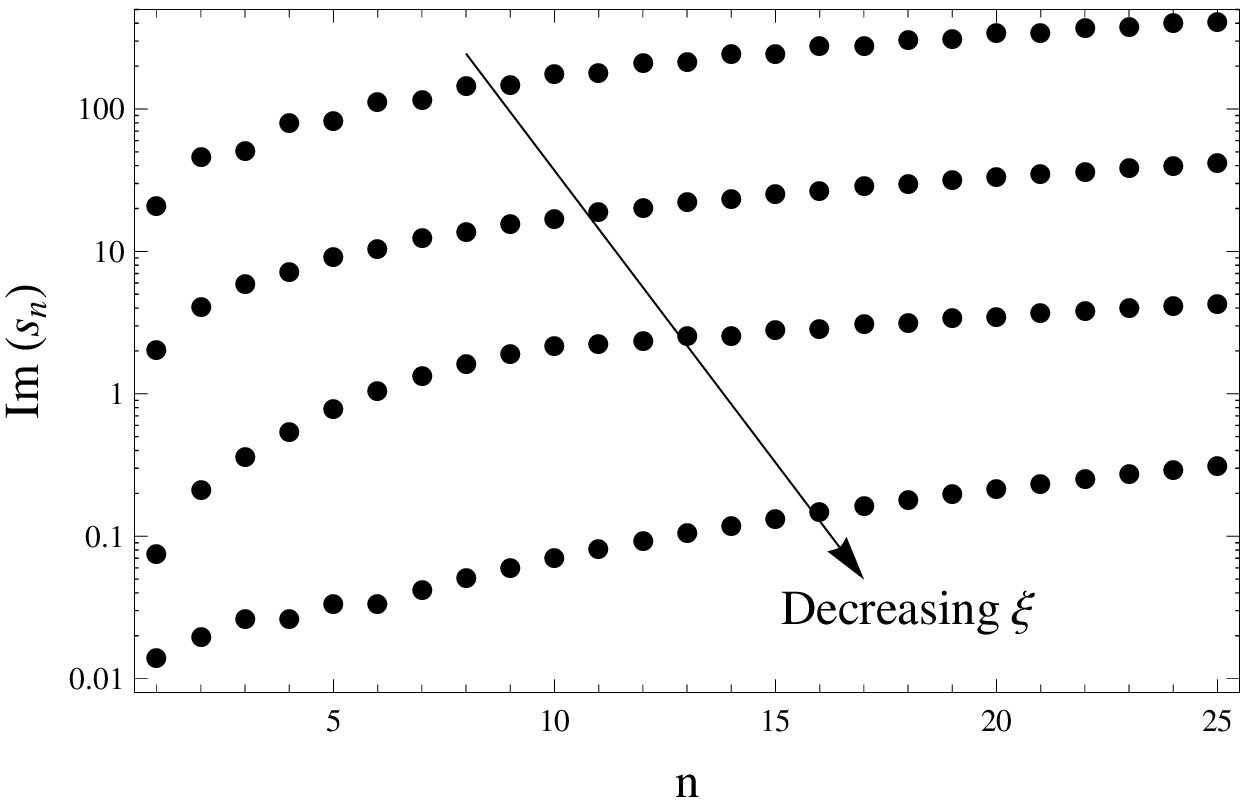} 
\caption{Left-hand panel: The RMS amplitude of f(t) as a function of $K$ for (top to bottom) $\xi=10$, $1$, $0.1$, $0.01$.  Right-hand panel: The imaginary component of the eigenvalues $s_n$ for $\xi=10$ (top row), $1$, $0.1$ and $0.01$ (bottom row). \label{fig5} }
\end{figure}

Before proceeding to neutron star parameters, we explore the effect of changing $K$ and $\xi$ on the oscillations.
In the left-hand panel of figure \ref{fig5}, the RMS amplitude of the oscillations is plotted as a function of $K$ for four values of $\xi$; from top to bottom: $10$, $1$, $0.1$ and $0.01$.  
We find that for all $\xi$, the oscillation amplitude vanishes as $K\rightarrow 0$, when the moment of inertia of the container is large compared with that of the fluid.  
The oscillation amplitude also vanishes in the opposite limit, as $K\rightarrow \infty$, because the angular momentum imparted to system when the container is spun-up is small.
The left-hand panel of figure \ref{fig5} also demonstrates that the RMS amplitude of oscillations also decreases as $\xi \rightarrow 0$.  This can be understood by examining the right-hand panel of figure \ref{fig5}, where the imaginary component of the eigenvalues $s_n$ are plotted for the same values of $\xi$ as the left-hand panel, namely $10$, $1$, $0.1$ and $0.01$ from top to bottom.
We find that the spacing between adjacent eigenvalues for a given $\xi$ is of the order of $\xi$; we have $s_{n+1}-s_n\sim 10$ and $0.01$ for the top and bottom rows, respectively. 
Interestingly, in the limit $\xi\rightarrow 0$, the eigenvalues become infinitely close and the spectrum becomes a continuum.
The response of an oscillating crust coupled to a continuum in the core has been studied in the context of magnetar quasi-periodic oscillations by \citet{lev06,lev07,van12}, who demonstrate that the resonant absorption of crust oscillations by the continuum results in the damping of crust oscillations in a fraction of an Alfv\'en crossing time (referred to as Landau damping).
The same effect is present here, except that the continuum arises because of the dependence of the spectrum on $\xi$ rather than the spherical geometry in the case of magnetar oscillations.
In the limit $\xi\rightarrow \infty$, studied by \citet{eas79a} and in \S\ref{secA2}, the continuum has no edges or turning points and the container oscillations are completely damped by this mechanism.
However, just as in the case of magnetar oscillations, Alfv\'en waves continue to propagate undamped in the core; the combined back-reaction of these waves on the crust conspire so that the resulting motion of the crust is zero.

\subsection{Application to neutron stars} \label{sec3b}

Having explored the general behavior of the solution, we now investigate the parameter space relevant to neutron stars.
The parameter $\xi$ is readily calculated for typical pulsar values
\begin{eqnarray} 
 \xi&=&1.7\times 10^{-4} \left(\frac{\it{B}}{10^{12}\, \rm{G}}\right) \left(\frac{\rho}{3.7 \times 10^{14} \, \rm{g\,cm^{-3}}}\right)^{-1/2}  \left(\frac{\it{R}}{1.2 \times 10^6 \, {\rm cm}}\right)^{-1} \left(\frac{P}{0.1\, \rm{s}}\right)\,, \label{eq42} \end{eqnarray}
where $R$ and $P$ are the pulsar radius and period respectively and we have evaluated (\ref{eq42}) at the average density of a typical pulsar.  
If the interior of the pulsar is a type II superconductor, the magnetic field is comprised of vortices with quantized magnetized flux plus the so-called London field, which arises from macroscopic vorticity.
The complete equations of motion in the superconducting and superfluid core of a neutron star have recently been derived by \citet{gla11}.
In the absence of a neutron fluid (and hence entrainment and mutual friction), the stress tensor for a type II superconductor can be extracted from the results of \citet{gla11}.
We obtain
\begin{equation}
 T^{SC}_{ij}=-p \delta_{ij}+ \frac{1}{4\pi} \left(b^L_i b^L_j -\frac{b^L_k b^L_k}{2} \delta_{ij}\right)+  \rho\nu_s  |\omega| \left( \hat{\omega}_i \hat{\omega}_j -\delta_{ij}\right) \,,\label{eq3a}
\end{equation}
where 
\begin{eqnarray}
  {\boldsymbol \omega}&=&\nabla \times {\boldsymbol v}+\frac{e}{mc}{\boldsymbol B}\,, \label{eq4e} 
\end{eqnarray}
is the vortex areal density multiplied by the circulation per vortex and
\begin{eqnarray}
 {\boldsymbol b}^L&=& -\frac{mc}{e}\nabla\times {\boldsymbol v}\,,
\end{eqnarray}
is the London field.
Equation (\ref{eq4e}) may also be cast into an equation for the total magnetic field, which is comprised of the vortex areal density multiplied by the magnetic flux per vortex  $m c {\boldsymbol \omega}/e$ and the London field, ${\boldsymbol b}^L$.
Equation (\ref{eq3a}) is comprised of contributions from the pressure, London field and the vortex line tension, respectively.
In (\ref{eq3a}), the vortex line tension parameter is defined
\begin{equation}
\nu_s=\frac{\hbar}{4 m}   \log\left(\frac{\Lambda}{\xi_p}\right)\,,
\end{equation}
where $\hbar$ is the reduced Planck constant, $m$ is the proton mass, $\xi_p$ is the coherence length for the proton condensate and 
\begin{equation}
\Lambda= \frac{1}{\sqrt{4 \pi \rho}} \left(\frac{m c}{e}\right) \,,
\end{equation}
is the London depth.
Evaluating the relative contribution of the London field in a neutron star, we find
\begin{equation}
  \lvert \frac{ {\boldsymbol b}^L}{{\boldsymbol B}} \rvert = 1.3\times 10^{-14} \left(\frac{P}{0.1 \, \rm{s}}\right)^{-1}  \left(\frac{ B }{10^{12} \, \rm{G}}\right)^{-1} \,.
\end{equation} 
Hence, the London field is a negligible contribution to the magnetic field, and (\ref{eq4e}) is well approximated by
\begin{eqnarray}
  {\boldsymbol \omega}&=&\frac{e}{mc}{\boldsymbol B}\,. \label{eq4c} 
\end{eqnarray}
The lower critical field for a type II superconductor is related to the tension parameter by 
\begin{equation}
H_{\rm cl}=\frac{ m c  \nu_s}{e \Lambda^2} \,,
\end{equation}
therefore the relative magnitudes of the second and third terms in (\ref{eq3a}) is
\begin{equation}
 \frac{ \left( b^L\right)^2}{ H_{\rm cl} |B|} = 4.2\times 10^{-31} \left(\frac{P}{ 0.1 \, \rm{s}}\right)^{-2}  \left(\frac{ B }{10^{12} \, \rm{G}}\right)^{-1} \left(\frac{ H_{\rm cl} }{4 \times 10^{14} \, \rm{G}}\right)^{-1} \,.
\end{equation} 
Hence, the London field contribution to the stress tensor is also negligible and (\ref{eq3a}) can be written 
\begin{equation}
 T^{SC}_{ij}=-p \delta_{ij}+ \frac{1}{ 4 \pi} H_{\rm cl}  |B| \left( \hat{\omega}_i \hat{\omega}_j -\delta_{ij}\right) \,,\label{eq3b}
\end{equation}
Equation (\ref{eq3b}) is the result obtained by \citet{eas77}, and is an extremely good approximation in a neutron star.
Replacing (\ref{eq3}) with (\ref{eq3b}), one finds that the analysis in \S\ref{sec2} follows identically, with the redefinition 
\begin{eqnarray}
\xi=\frac{1}{L\Omega}\sqrt{\frac{B_0 H_{\rm cl}}{4 \pi \rho}}\,, \label{eq16b}
\end{eqnarray}
as originally predicted by \citet{eas79a} and used by subsequent authors \citep{men98,and09}.
For a type II superconductor we therefore have 
\begin{eqnarray} 
 \xi&=&4.7\times 10^{-3} \left(\frac{\it{B}}{10^{12}\, \rm{G}}\right)^{1/2}\left(\frac{H_{\rm cl}}{4\times 10^{14}\, \rm{G}}\right)^{1/2} \left(\frac{\rho}{3.7 \times 10^{14} \, \rm{g\,cm^{-3}}}\right)^{-1/2}  \nonumber \\
&& \times \left(\frac{\it{R}}{1.2 \times 10^6 \, {\rm cm}}\right)^{-1} \left(\frac{P}{0.1\, \rm{s}}\right)\,. \label{eq42b} 
\end{eqnarray}

To estimate the Ekman number we require the viscosity in the neutron star outer core.
One typically uses the expression for electron-electron scattering obtained by \citet{cut87}, who approximately evaluated the calculations of \citet{flo76}  using the tables in \citet{bay71} for non-superfluid nuclear matter:
\begin{equation} \label{eq42a}
 \eta_{ee}=6.0\times10^{18}\, \rm{g\,cm^{-1} s^{-1} }\left(\frac{\rho}{10^{14} \, \rm{g\,cm^{-3}}}\right)^2\left(\frac{\it{T}}{10^8 \, \rm{K}}\right)^{-2}\,.
\end{equation}
where $T$ is the temperature.
This result is valid over the density range $10^{14}<\rho<4\times 10^{14}$.
More recent calculations for the electron viscosity have been performed by \cite{and05} and \citet{sht08} accounting for superfluidity of proton matter.  The former finds results broadly consistent with (\ref{eq42a}), while the latter accounts for transverse Landau damping in charged particle collisions neglected in previous studies, an effect that results in viscosities a factor of three smaller than that predicted by (\ref{eq42a}).  \citet{sht08} also demonstrate that the electron viscosity is over an order of magnitude larger than the neutron viscosity when the protons are superfluid, making it the dominant contribution in neutron star cores.  Using (\ref{eq42a}) divided by three, the Ekman number in a neutron star is approximately
\begin{equation} \label{eq43}
 E=7.3\times 10^{-10} \left(\frac{\rho}{3.7 \times 10^{14} \, \rm{g\,cm^{-3}}}\right) \left(\frac{\it{T}}{10^8\, \rm{K}}\right)^{-2} \left(\frac{\it{R}}{1.2 \times 10^6 \, {\rm cm}}\right)^{-2} \left(\frac{P}{0.1\, \rm{s}}\right)\,.
\end{equation}
Equation (\ref{eq43}) is sensitive to the temperature; an order of magnitude reduction in the temperature results in a factor of a hundred increase in the Ekman number, so that viscosity becomes much more important in older pulsars.

To estimate $K$, we use the canonical value for the (inverse) inertia fraction of the crust, $K\approx 50$ \citep{sha83}.  
This number is also broadly consistent with the lower bound obtained from first principles calculations by \citet{lat07}, which is also discussed in \citet{van10}.

To estimate $\alpha$, we use the classical expression for the torque exerted by electromagnetic dipole radiation \citep{jac98}, namely $ \tau_{ext}=(2/3) B^2 R^6 \sin^2 \theta c^{-3} $ (in cgs units), where $\theta$ is the angle between the magnetic and rotation axes, and $c$ is the speed of light.  Taking $\sin\theta\sim 1$ and assuming the pulsar is a rigidly rotating sphere with uniform density, we obtain
\begin{eqnarray} 
\alpha&=&9.7\times 10^{-15} \left(\frac{\it{B}}{10^{12}\, \rm{G}}\right)^2 \left(\frac{\rho}{3.7 \times 10^{14} \, \rm{g\,cm^{-3}}}\right)^{-1} \nonumber \\
&& \times \left(\frac{\it{R}}{1.2 \times 10^6 \, {\rm cm}}\right) \left(\frac{P}{0.1\, \rm{s}}\right)^{-2}  \left(\frac{\epsilon}{10^{-6}}\right)\,. \label{eq43a}
\end{eqnarray}

\begin{figure}
\epsscale{0.40}
\plotone{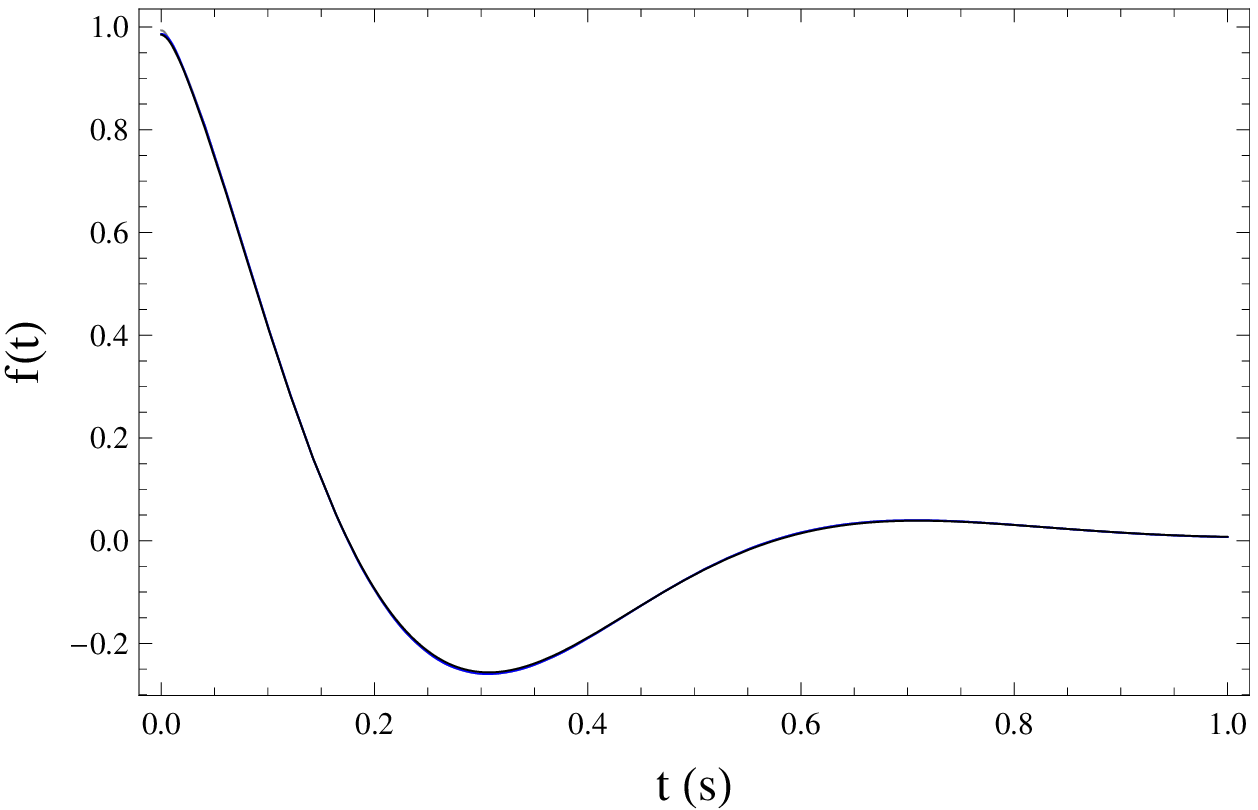} 
\plotone{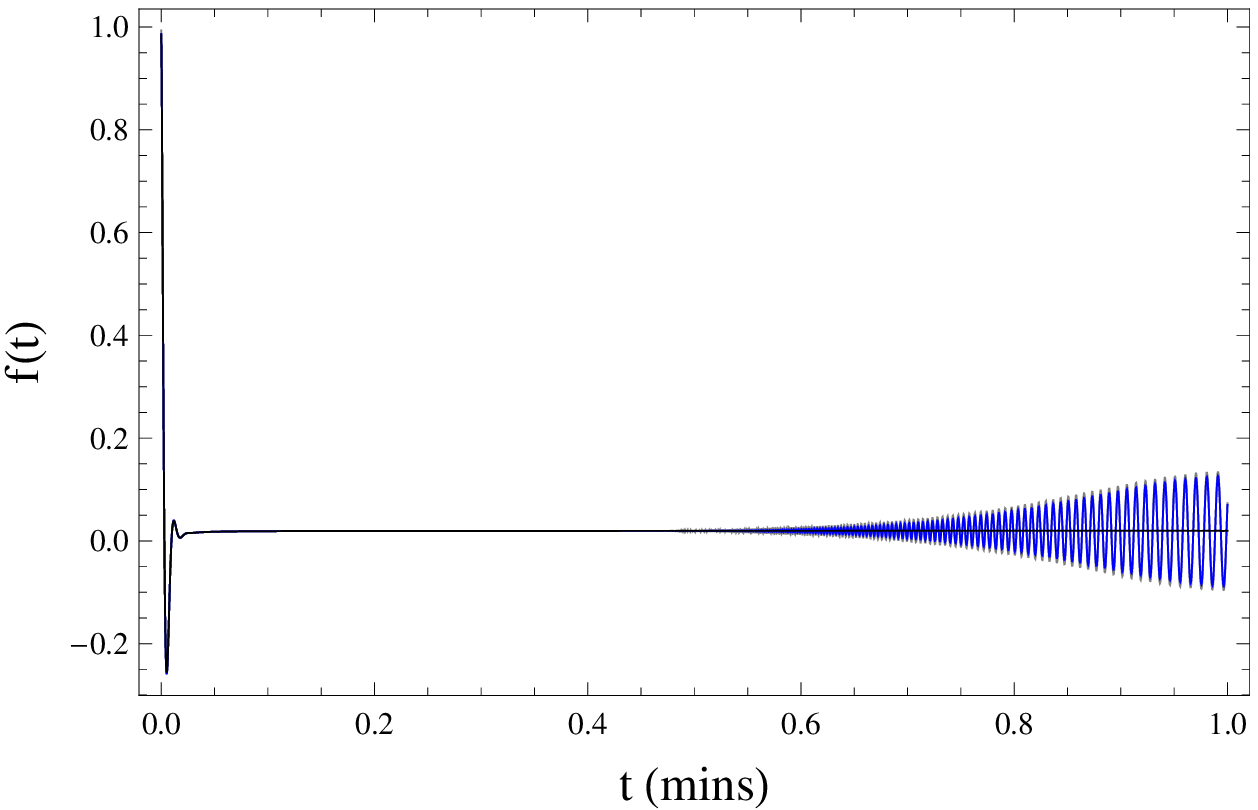} \\
\plotone{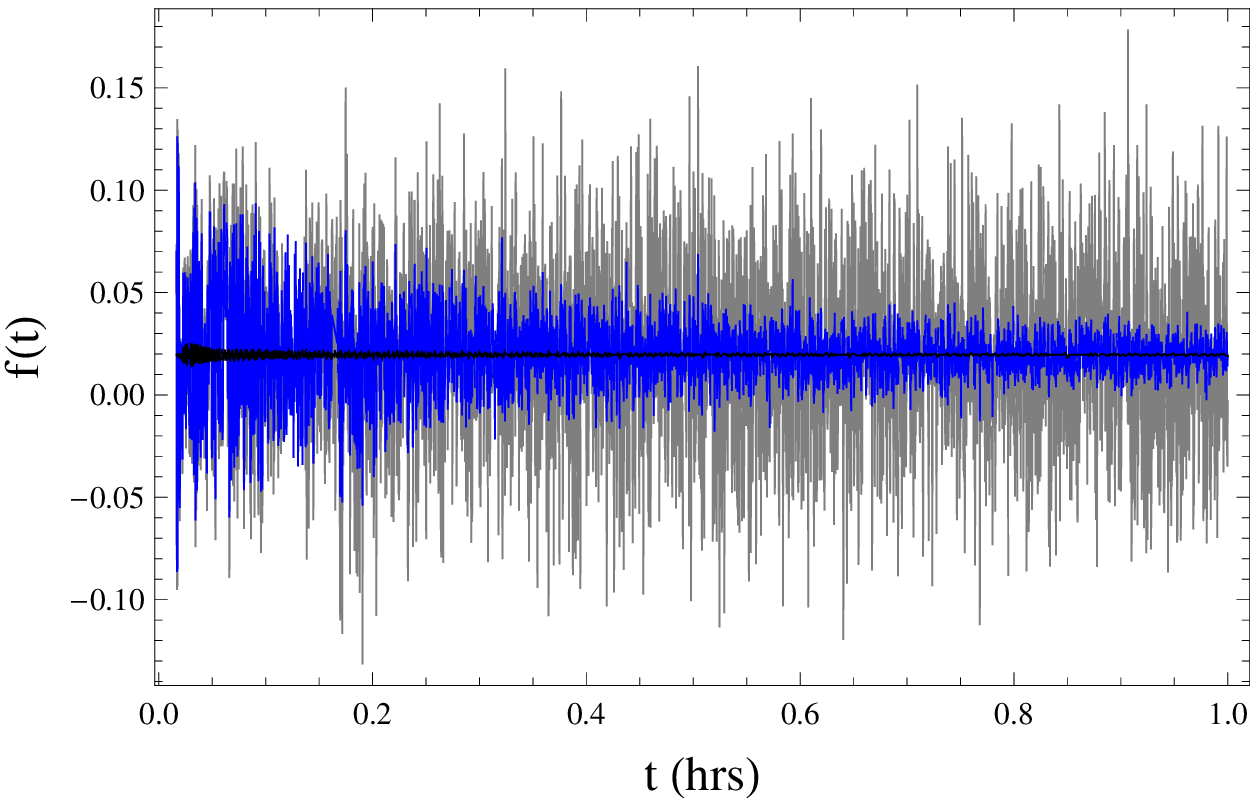} 
\plotone{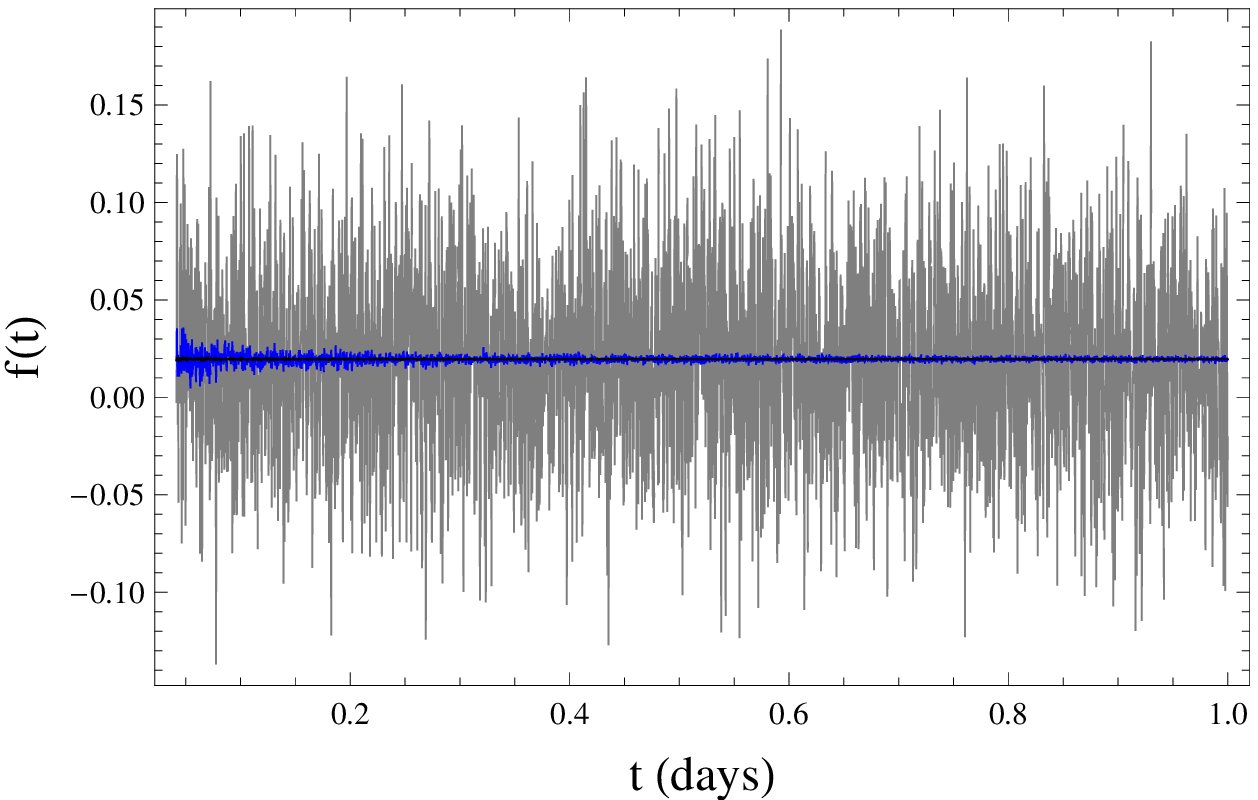} \\
\caption{The response of the crust following a glitch, $f(t)$, assuming $\xi=10^{-3}$, $K=50$, $\alpha=10^{-14}$, $\Omega=70.3\,{\rm rad\, s^{-1}}$  and $E=0$ (gray curve), $E=10^{-9}$ (blue curve), $E=10^{-7}$ (black curve).  The response is plotted for the first second (top-left), first minute (top-right), from one minute to one hour (bottom-left) and one hour to one day (bottom-right).\label{fig6} }
\end{figure}

In figure \ref{fig6}, we plot the general solution (\ref{eq39}) as a function of time for the Vela pulsar, taking $\xi=10^{-3}$, $K=50$, $\alpha=10^{-14}$ and $\Omega=70.3\,{\rm rad\,s^{-1}}$.
Dimensional time is used for easy comparison with glitch recovery time-scales.
To compare different viscosities, curves for three values of the Ekman number are plotted, $E=0$ (gray curve), $E=10^{-9}$ (blue curve) and $E=10^{-7}$ (black curve).  
In the top panels, we plot the response of the pulsar crust $f(t)$ during the first second (left-hand panel) and first minute (right-hand panel) following a glitch.
We find that the initial response of the crust is the same for all values of the viscosity.
Oscillations begin after an Alfv\'en crossing time ($\sim 30 \,{\rm sec}$) and grow to the same amplitude ($\sim 10\%$ of the glitch amplitude) for the gray ($E=0$) and blue ($E=10^{-9}$) curves, but are significantly suppressed for the black ($E=10^{-7}$).
In the lower panels we plot the response at later times; the following hour in the left hand panel, and the following day in the right-hand panel.
The oscillations are damped by viscosity to a factor of $1/e$ of their maximum amplitude over a timescale of 15 minutes for the black  curve ($E=10^{-7}$), and 30 minutes months for the blue curve ($E=10^{-9}$).
Interestingly, these numbers are of similar magnitude and are greater than the Ekman times ($44\, {\rm sec}$ and $7.5\,{\rm mins}$ respectively) and much less than the corresponding diffusion times ($39\, {\rm hrs}$ and $5.3\,{\rm months}$ respectively).

\begin{figure}
\epsscale{0.40}
\plotone{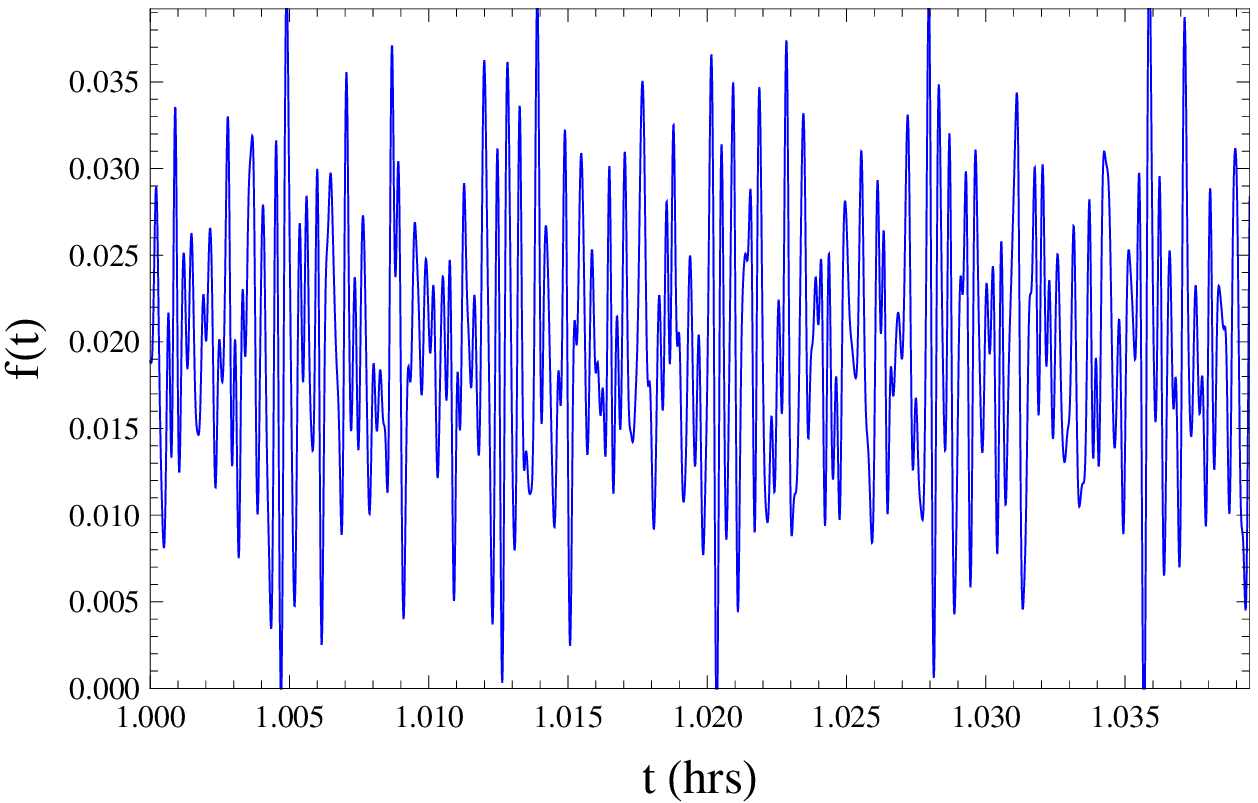} 
\plotone{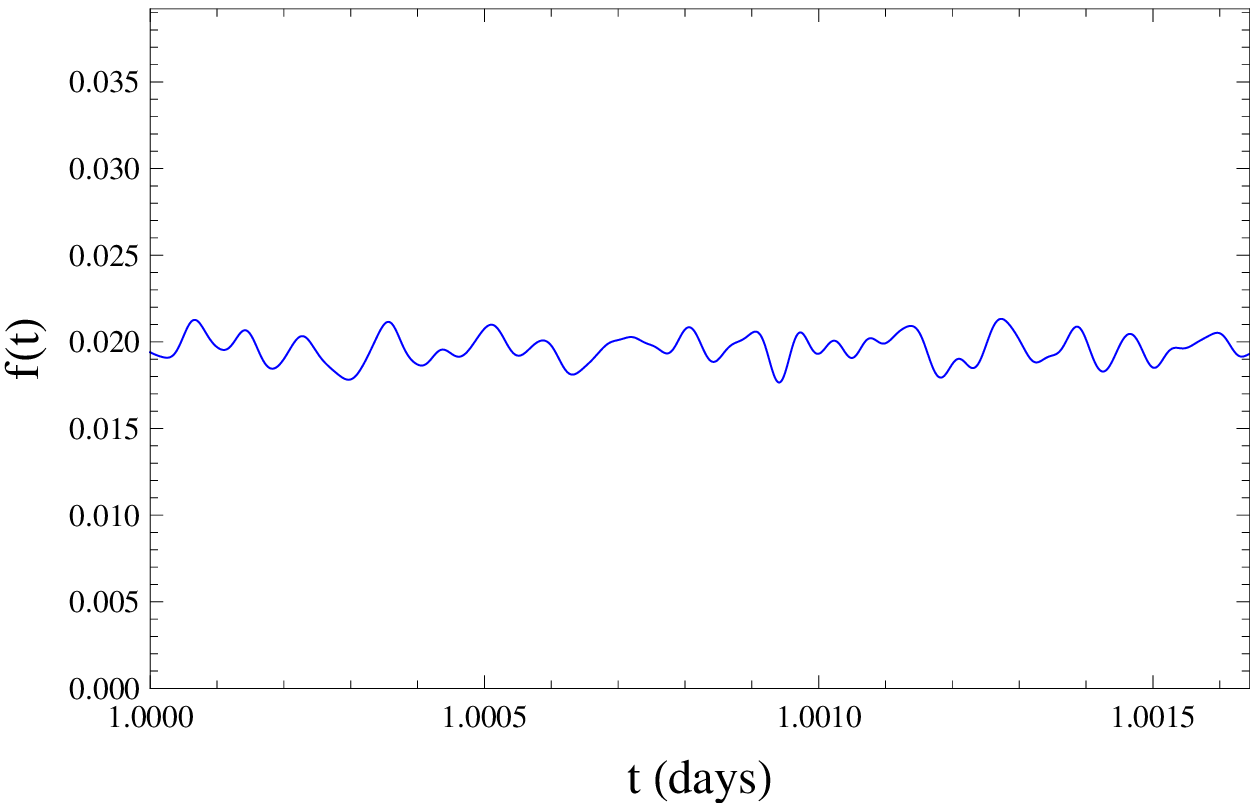} \\
\plotone{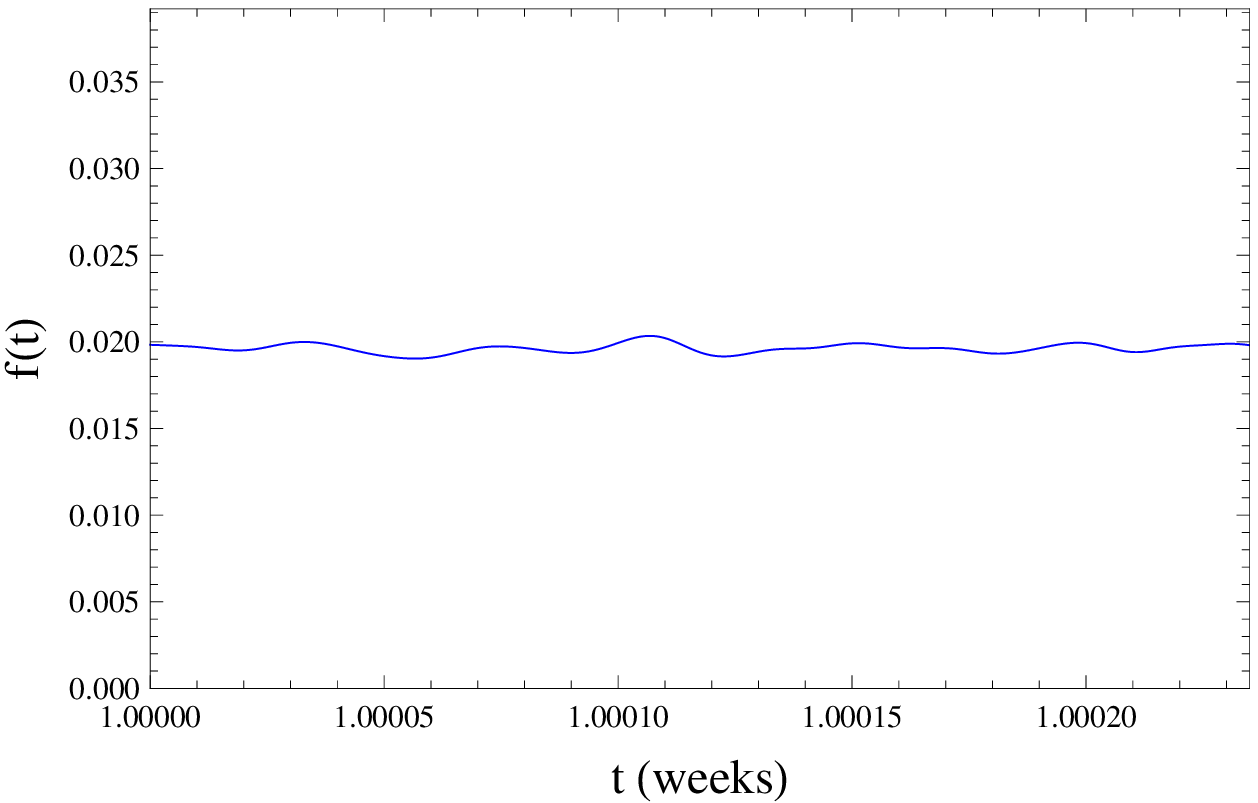} 
\plotone{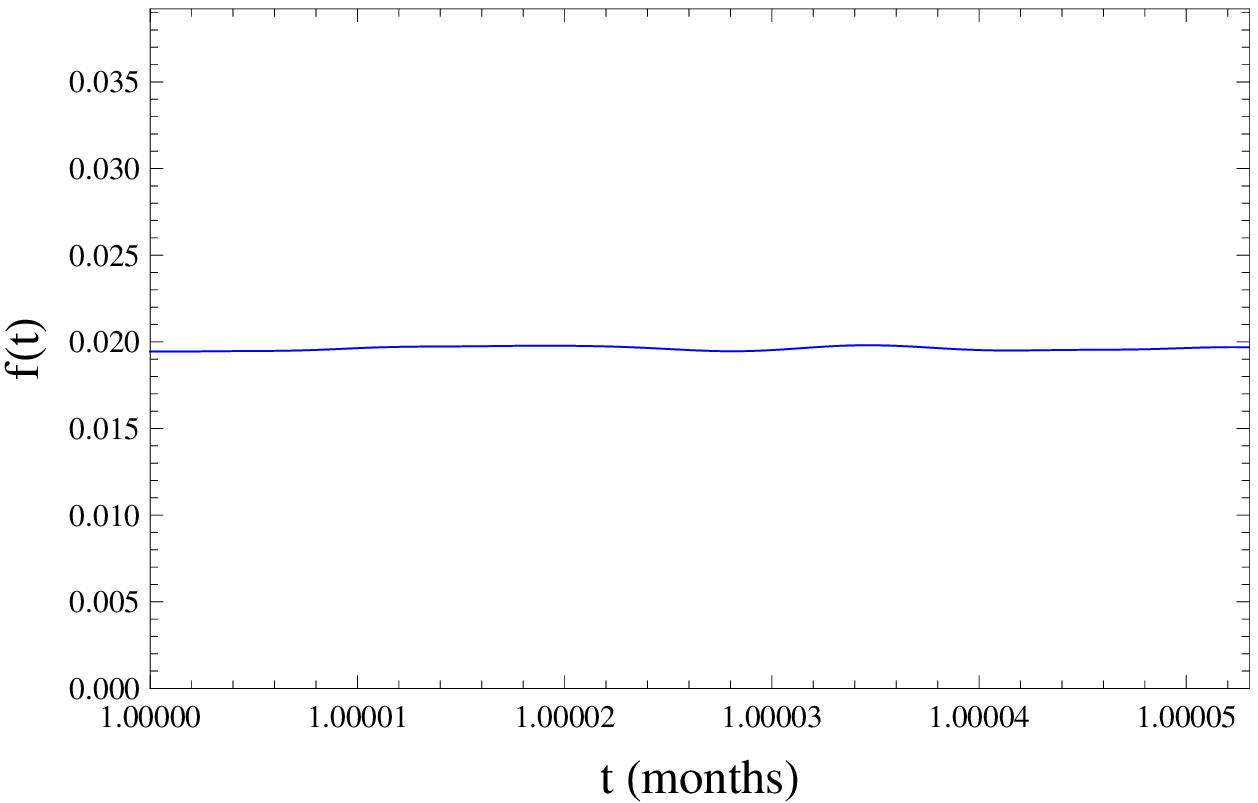} \\
\caption{The response of the crust following a glitch, $f(t)$, for $\xi=10^{-3}$, $K=50$, $\alpha=10^{-14}$, $E=10^{-9}$ and $\Omega=70.3\,{\rm rad\,s^{-1}}$.  Each panel shows a period of $2.5\,{\rm mins}$, corresponding to five Alfv\'en crossing times and has the same vertical scale for comparison.  The different panels correspond to one hour (top-left), one day (top-right), one week (bottom-left) and one month (bottom-right) after the glitch.  \label{fig7} }
\end{figure}

Because of the different scales present in the response, it is difficult to ascertain many aspects of the oscillations from figure \ref{fig6}.  
In figure \ref{fig7}, the response for $E=10^{-9}$ is plotted for a period of five Alfv\'en crossing times ($\sim2.5\,{\rm mins}$) at four different times after the glitch: an hour (top-left), a day (top-right), a week (bottom-left) and a month (bottom-right).
At early times, the crust oscillates rapidly, with a period of less than a second.
At later times, the amplitude and frequency of the oscillation decreases as the higher wavenumber magneto-inertial modes propagating inside the star are damped more rapidly by viscous forces.
After a day (week) the oscillation period has increased to roughly $8\,{\rm s}$ ($25\,{\rm s}$), and the amplitude is $0.3\%$   ($0.1\%$) of the glitch amplitude.

\begin{figure}
\epsscale{0.40}
\plotone{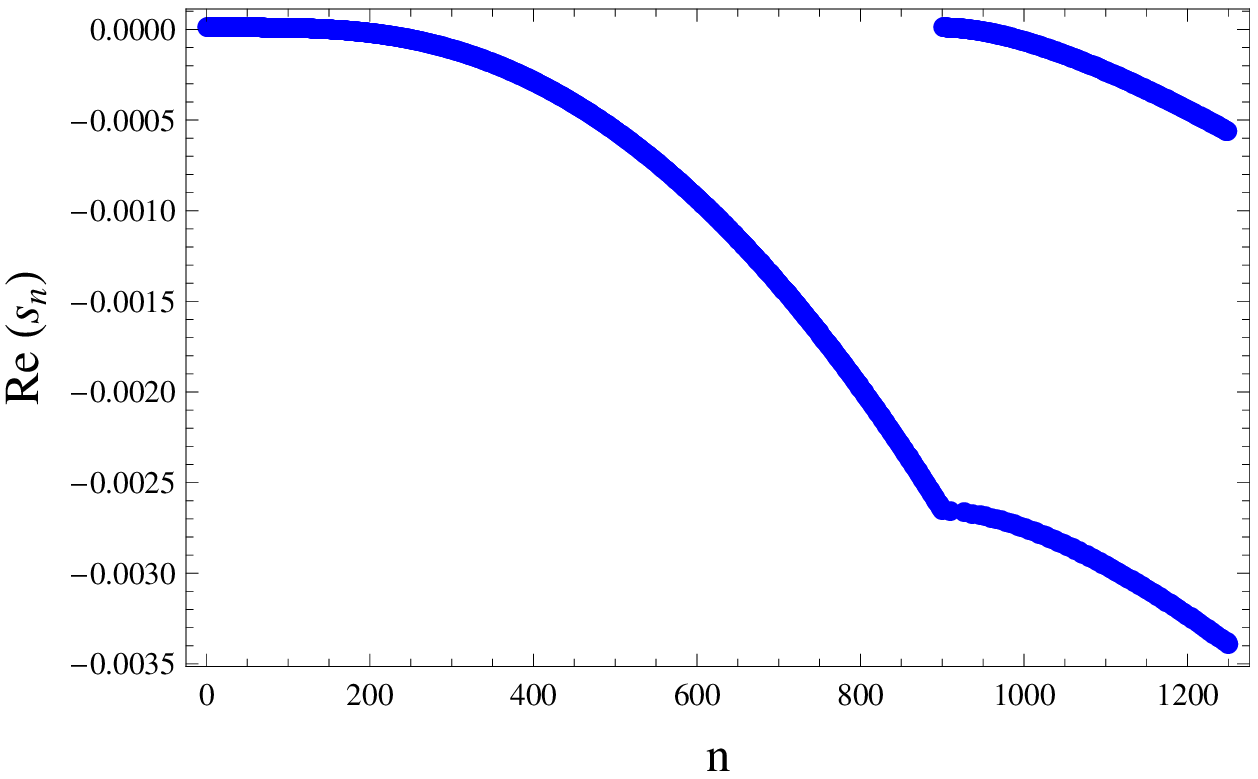} 
\plotone{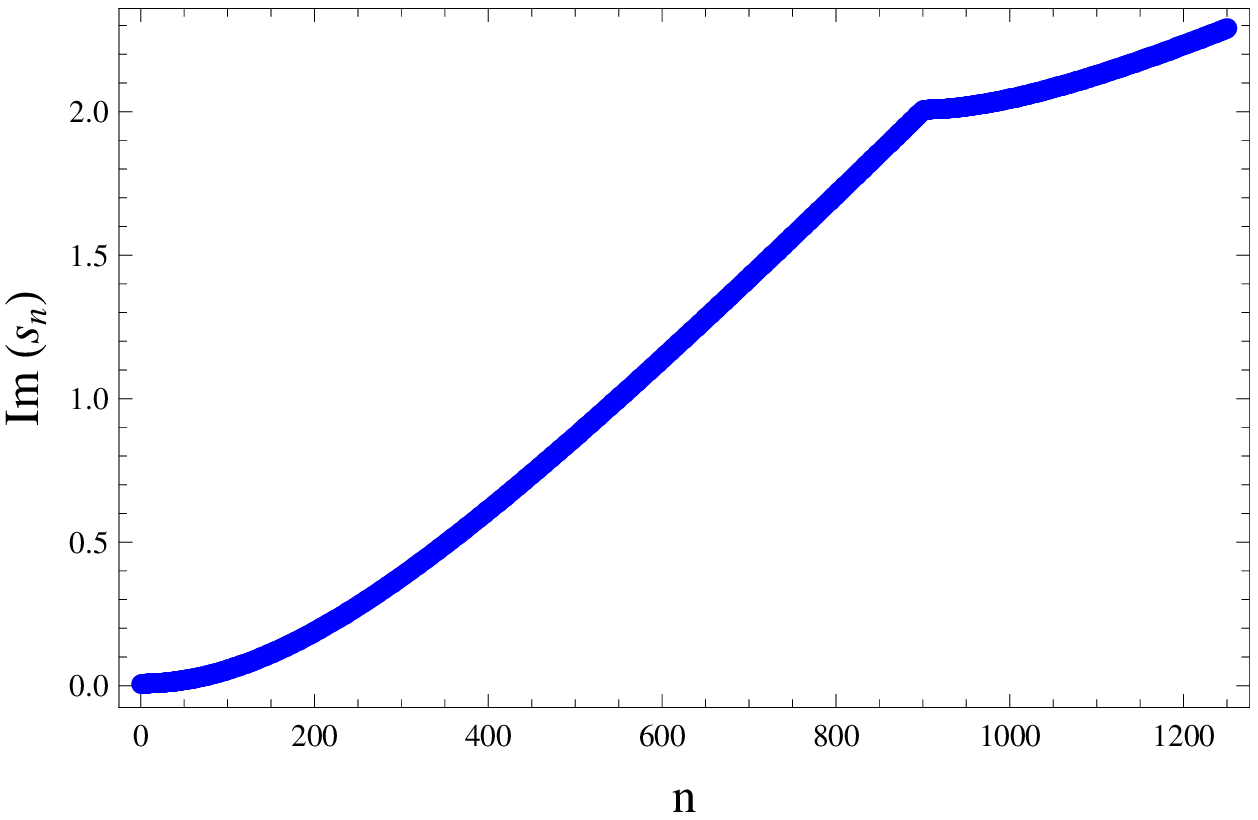} \\
\plotone{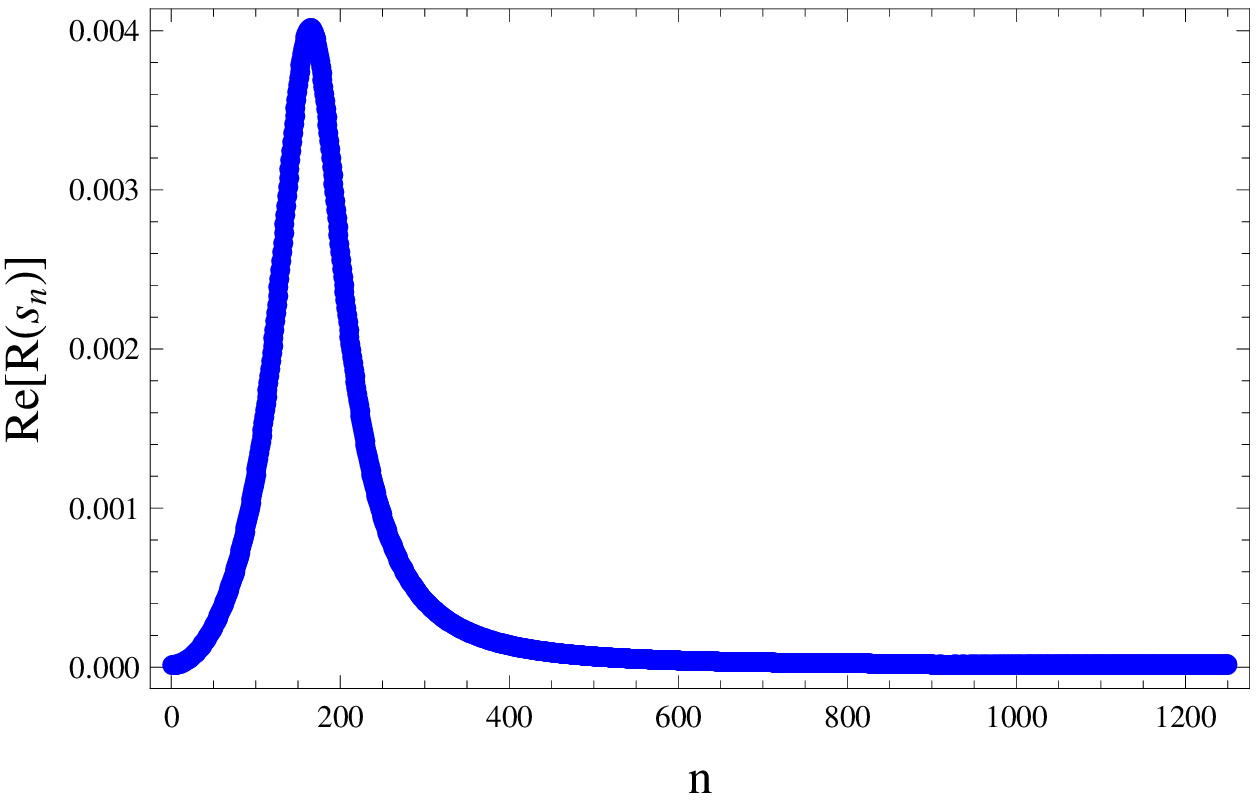} 
\plotone{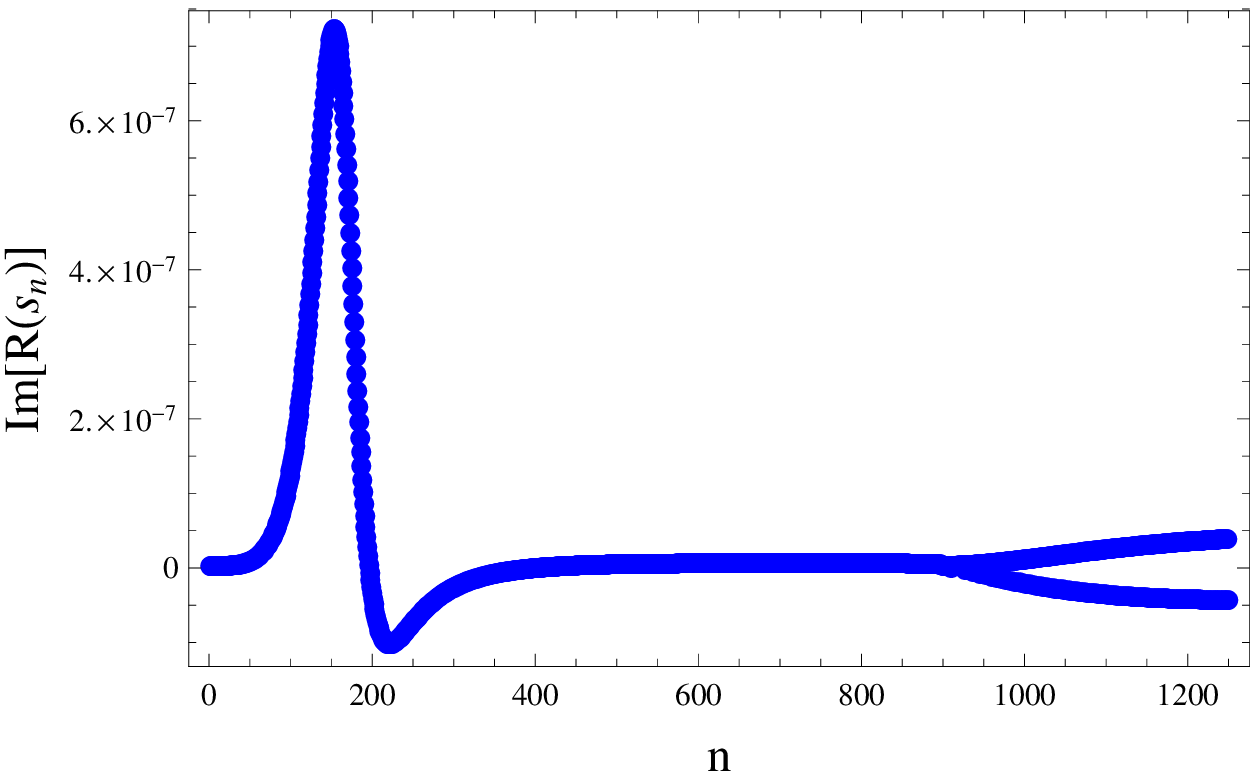} \\
\caption{Top panels: the real (left) and imaginary (right) components of the eigenvalues $s_n$ for the $\xi=10^{-3}$, $K=50$ and $E=10^{-9}$ solution plotted in figures \ref{fig6} and \ref{fig7}.   \label{fig8} }
\end{figure}

To investigate the aspects of the $E=10^{-9}$ solution further, in figure \ref{fig8} we plot the real (top-left) and imaginary (top-right) components of $s_n$, and the real (bottom-left) and imaginary (bottom-right) components of $R(s_n)$.  The real and imaginary components of $s_n$ relate directly to the damping time $[ {\rm Re}(s_n) \Omega]^{-1}$ and period [$ 2 \pi /{\rm Im}(s_n) \Omega$] of the modes in (\ref{eq39}), while  $ 2 R(s_n)$ corresponds to the initial amplitude of an oscillation mode as a fraction of the initial glitch amplitude.
Examining the bottom-right panel, we find that the dominant mode occurs at $n=165$, for which $s_{165}=-1.5 \times 10^{-5}+0.13 i$ and $R(s_{165})=4.0 \times 10^{-3}+ 6.2 \times 10^{-7}i$.  
This mode has a period of $0.7\,{\rm s}$, and is damped out over a short timescale of $15.8\,{\rm min}$.  
The initial amplitude is $4.0 \times 10^{-3}$ times the glitch amplitude.
Therefore, this mode oscillates too rapidly and is too short-lived to be resolvable by radio telescopes.
The mode with the longest period is the first mode, which has $s_1=-5.9\times 10^{-14}+7.7\times 10^{-6} i$ and$R(s_{1})=8.2\times 10^{-8}+1.3 \times 10^{-15} i $.
The corresponding period is $3\,{\rm hrs}$ with a damping time of $7.6 \times 10^3\,{\rm yrs}$.
However its initial amplitude is only $1.6 \times 10^{-7}$ times the glitch amplitude and therefore too small to be observed.
The most likely candidates for detection lie around $n=8$, which has a period of $4.5\,{\rm mins}$ and a damping time of $4\,{\rm yrs}$. The initial amplitude is $1.2 \times 10^{-5}$ times the glitch amplitude, approaching detectability for a glitch of magnitude $10^{-6}$, however the period is too short to be resolved by radio timing data.
Therefore, for modes with an amplitude sufficiently large for detection, the period and damping time of the oscillations are too short to be resolvable by radio telescopes. 

A number of interesting phenomena are observed in pulsars that exhibit quasi-periodic oscillations.
The first is following the 1988 Christmas glitch in Vela, where a damped periodic oscillation is clearly visible in the timing residuals [see figure 2 of \citet{mcc90}]. However, these oscillations appear to have a period of $\sim 20\, {\rm days}$, far too long to be explained by the oscillations predicted in this paper.
The second is the ``overshoot'' observed in the glitch recovery of the Crab, present in the 1975 and 1986 glitches \citep{won01,van10}.  This could potentially be explained as a damped oscillation, however, in this case the oscillation period would correspond to weeks, again, too long to correspond to the oscillations here.
Another possibility is that the irregular and quasi-periodic nature of the oscillations predicted in this paper manifest themselves as some form of timing noise.  Again, however, we find that the oscillations predicted here are way too fast and are damped far too quickly; quasi-periodic oscillations observed in timing noise have periods of the order of years \citep{hob10,lyn10}. 

The results in this paper have been obtained using a toy model based on cylindrical geometry with a uniform magnetic field, and may change for more realistic neutron star models.
As discussed in \S\ref{sec3a}, in a non-rotating star spherical and cylindrical geometry produce qualitatively different results.
In the former case, the Alfv\'en spectrum in the core comprises a continuum, resulting in the Landau damping of crust oscillations, an effect that is not captured by the cylindrical geometry considered here.
Persistent oscillation modes of the crust then correspond to frequencies at the turning points and edges of the continuum, and crustal modes that lie in the gaps of the continuum \citep{lev07,van12}.
However, the generalization to a rotating star is not obvious.  
The solution to (\ref{eq15}) in spherical geometry is non-trivial, and rotation may destroy the continuum like factors such as magnetic field geometry, superfluidity and stratification have been shown to do in non-rotating stars \citep{col09,gab13,pas14}.
A realistic prediction of the quasi-period oscillations following a glitch requires a more detailed study than that presented here, and most likely treated numerically.

In a non-magnetized star, stratification has been shown restrict the penetration depth of Ekman pumping, shortening the spin-up time and decoupling the core of the star \citep{abn96,van09}.
An analogous result is expected in a magnetized plasma \citep{men98}, although a complete calculation of the spin up of a uniformly magnetized stratified plasma has not yet been performed \citep{mel12}.
However, just as in the present study, magneto-inertial oscillations should persist in the core after the spin-up has completed.
These waves will be affected by stratification, which may alter the amplitude, period and onset of the oscillations predicted in this paper.

Realistic magnetic field geometries (including poloidal and toroidal components) and misalignment of the magnetic axis with the rotation axis also add another level of complexity, and it is unclear how these factors will affect the present results.
Changes in the angular velocity of the crust may not be communicated to regions of the star containing closed magnetic field lines, which may decouple, reducing the effective moment of inertia of the star \citep{eas79c}.
In a stratified star with realistic field geometries the problem becomes significantly more difficult, e.g. regions of the star where the phase velocity of magneto-inertial waves vanishes may decouple; the reader is referred to \citet{mel12} for a comprehensive discussion on the topic.

The neutron superfluid has also been neglected in this paper.  
This component comprises the bulk of the fluid in the outer core and couples to the plasma via a mutual friction force \citep{men91b,gla11}, which arises from the interaction of magnetized neutron vortices with electrons \citep{alp84} or type II superconducting flux tubes \citep{rud98}.
The multi-fluid hydrodynamics gives rise to a much richer spectrum of oscillation modes that have been studied in the literature, and have recently been studied numerically in the context of glitches by \citet{pas11}, who assumed that the crust and proton-electron plasma were locked together by the magnetic field.
It would be interesting to see this combined with crust oscillations and magnetic fields to obtain a complete picture of neutron star seismology.  
The effects discussed in this paper (e.g., Ekman pumping and Landau damping) should manifest themselves in such a study.

\section{Conclusions} \label{sec4}

We have shown that co-rotation of the crust and plasma in the outer core of the pulsar following a glitch is not brought about by the magnetic field.  Rather, the glitch excites magneto-inertial waves that propagate through the plasma, generating torsional oscillation modes of the crust as they reflect internally.
The oscillations have a period of $1-20\,{\rm s}$ and decay over a timescale of $15-30\, {\rm mins}$ due to the viscosity of the electrons in the core.

The toy model presented here is unlikely to represent a realistic oscillation spectrum following a glitch, however, it sheds light on some qualitative features of the problem.
First, co-rotation between the plasma and crust of a neutron star cannot be achieved by the magnetic field following a glitch.  This is a violation of conservation of energy; the system oscillates persistently until it is damped by electron viscosity.
Second, an Ekman pumping mechanism, first identified by \citet{eas79a} is present at short times, which spins up the plasma in the interior.
However, following the spin-up, Alf\'en waves excited by the glitch continue to propagate through the plasma, exciting oscillations of the crust after an Alfv\'en crossing time.
Third, for rapidly rotating stars (where rotation energy dominates the magnetic energy, i.e., $\xi \rightarrow 0$), the spectrum of magneto-inertial oscillations in the core becomes a continuum and the oscillations of the container undergo Landau damping.  
This in analogous to the resonant absorption mechanism identified in magnetars, where the oscillation modes in the core conspire so that the net back-reaction on the crust vanishes.

To determine a realistic spectrum for a neutron star following a glitch the use of a numerical code such as those used to model magnetar quasi-period oscillations is most likely required.
These codes have achieved success in reproducing the observed QPO frequencies for non-rotating stars using relativistic ideal MHD, and have recently incorporated the essential physics of superfluidity, superconductivity and stratification, although these studies are still in their infancy.
It would be interesting to incorporate magnetic fields into models for rotating stars and glitch recovery, as it is possible that there is physics occurring that has not been resolved by existing radio telescopes, but which may be observable by future radio telescope arrays such as LOFAR and the SKA, gravitational wave detectors and x-ray observatories.
The discovery and identification of such a spectrum following a glitch would be extremely useful for constraining the physics of neutron star interiors.

\acknowledgments

Thanks to Matthias Rheinhardt, Bennett Link, Andrew Melatos and Axel Brandenburg for their useful feedback on the manuscript.  

\appendix
\section{Limits}

In the Appendices we present analytical solutions to the inverse Laplace transform of (\ref{eq33}) in two limits: slow rotation and fast rotation. 
These results, for which elegant mathematical expressions can be obtained, are used to verify the more general solution (\ref{eq39}) in \S\ref{sec3a}.
Results are obtained for $\alpha=0$, but can easily be generalized to the case when $\alpha \neq 0$, as for (\ref{eq39}).

\subsection{Slow rotation limit} \label{secA1}

In the slow rotation limit, $\Omega\approx 0$ and we have $E \gg 1$, $\xi \gg 1$ and we take $\alpha=0$.
Strictly speaking, $\Omega$ disappears entirely from the equations, and is no longer appropriate for defining scaled variables and dimensionless quantities.
However, in order to compare results to those obtained from (\ref{eq39}), we write the solution in terms of the scaled quantities, noting that factors of $\Omega$ cancel out when writing in terms of dimensional variables.
In this limit we find
\begin{eqnarray}
  k_+&\approx &k_-\approx \frac{|s|}{\sqrt{E s +\xi^2}}\,, 
\end{eqnarray}
and hence
\begin{eqnarray}
\tilde{f}&=&  \frac{  k \cosh k  }{s\left(k   \cosh k + K \sinh k  \right) }  \,.
\end{eqnarray}
The poles of $\tilde{f}$ are at $s=0$ and at $k  \cosh k + K \sinh k =0$.
The inverse Laplace transform is
\begin{equation}
  f(t)=\frac{1}{1+K}+\sum_n \frac{2 K}{K\left(1+K\right)+\lambda_n^2} \left[ \left(\frac{E s_{n+} + \xi^2}{E s_{n+} +2 \xi^2}\right) e^{s_{n+} t}+  \left(\frac{E s_{n-} + \xi^2}{E s_{n-} +2 \xi^2}\right) e^{s_{n-} t}\right]\,. \label{eqA3}
\end{equation}
where $\lambda_n$ are the positive solutions of 
\begin{equation}
  \lambda_n  + K \tan \lambda_n =0 \,. \label{eqA4}
\end{equation}
and 
\begin{equation}
  s_{n\pm}=-\frac{E\lambda_n^2}{2}\pm \sqrt{ \left(\frac{E\lambda_n^2}{2}\right)^2-\lambda_n^2 \xi^2} \,. \label{eqA5}
\end{equation}
Because $\Omega=0$, the viscous diffusion time and Alfv\'en crossing time are the only scales in the problem, and we can only consider the ratio of $E$ and $\xi$.
For $\xi> E\lambda_n/2$, $s_{n\pm}$ are complex conjugates.  For $E\ll \xi$, $s_{n\pm}=\pm i \lambda_n  \xi$, and the system is purely oscillatory. For $E\gg \xi$, $s_{n-}=-E \lambda_n^2$, the pre-factor of $e^{s_{n+} t}$ vanishes because $s_{n+}=0$ and the system relaxes exponentially.  
In this limit, the solution only applies to containers of infinite aspect ratio, as boundary condition for the azimuthal component of the velocity at a vertical boundary is not satisfied.

\subsection{Fast rotation limit} \label{secA2}

For fast rotation we consider the limit $E\ll 1$, $\xi \ll 1$, taking $\alpha=0$, 
We look for long time-scale solutions corresponding to Ekman pumping, and therefore assume $s\gg1$.
In this limit, $k_\pm \gg 1$ and hence $\cosh k_\pm \approx \sinh k_\pm \approx e^{k_\pm}/2$.  
Equation (\ref{eq33}) can be written as
\begin{equation}
\tilde{f}=\frac{s\left[2 \left(k_+k_-\right)-\left(k_++k_-\right)\right]+2 i  \left(k_--k_+\right)}{s^2\left[2 \left(k_+k_--K\right) -\left(k_++k_-\right)\left(1-K\right)\right]+2 i s \left(k_--k_+\right)\left(1+K\right)} \label{eqA6}
\end{equation}
Applying the limits above, we have
\begin{equation}
 k_\pm\approx\sqrt{\frac{ \pm 2 i s}{E s+\xi^2}}=\sqrt{\frac{ s}{E s+\xi^2}}\left(1\pm i\right)\,,
\end{equation}
and to leading order (\ref{eqA6}) becomes
\begin{equation}
\tilde{f}=\frac{s^{3/2}+\sqrt{E s+\xi^2}}{s\left[s^{3/2}+\sqrt{E s+\xi^2}\left(1+K\right)\right]} \,. \label{eqA8}
\end{equation}
Equation (\ref{eqA8}) can only be inverted for either $\xi=0$ and $E=0$.
For $\xi=0$, (\ref{eqA8}) reduces to
\begin{equation}
\tilde{f}=\frac{s+\sqrt{E}}{s\left[s+\sqrt{E}\left(1+K\right)\right]} \,, \label{eqA9}
\end{equation}
which has poles at $s=0$ and $s=-\sqrt{E}(1+K)$. 
The inverse Laplace transform is
\begin{equation}
 f(t)=\frac{1}{1+K}\left[1+K e^{-\sqrt{E}\left(1+K\right) t} \right]\,. \label{eqA10}
\end{equation}
For $E=0$, (\ref{eqA8}) reduces to
\begin{equation}
\tilde{f}=\frac{s^{3/2}+\xi}{s\left[s^{3/2}+\xi \left(1+K\right)\right]} \,.
\end{equation}
which has poles at $s=0$, $s=1$ and $s=-(1\pm i \sqrt{3})\xi^{2/3}(1+K)^{2/3}/2$. 
The inverse Laplace transform is
\begin{eqnarray}
 f(t)&=&\frac{1}{3 \left(1+K\right)}\left\{3+K e^{\xi^{2/3} \left(1+K\right)^{2/3} t}  {\rm Erfc}\left[\xi^{1/3} \left(1+K\right)^{1/3} \sqrt{t}\right] \right. \nonumber \\
&+& K e^{-\frac{1}{2} \left(1-i \sqrt{3}\right) \xi^{2/3} \left(1+K\right)^{2/3} t}   {\rm Erfc}\left[-\frac{1}{2} \left(1+i \sqrt{3}\right) \xi^{1/3} \left(1+K\right)^{1/3} \sqrt{t}\right] \nonumber \\
 &+& \left. K e^{-\frac{1}{2} \left(1+i \sqrt{3}\right) \xi^{2/3} \left(1+K\right)^{2/3} t} {\rm Erfc}\left[-\frac{1}{2} \left(1-i \sqrt{3}\right) \xi^{1/3} \left(1+K\right)^{1/3} \sqrt{t}\right]\right\} \label{eqA12}
\end{eqnarray}
where ${\rm Erfc}(x)$ denotes the complementary error function.

\bibliographystyle{avebib}
\bibliography{references}

\end{document}